\documentclass[11pt, reqno,preprint]{article}
\usepackage{jheppub}
\usepackage{epsfig}
\usepackage{amssymb}
\usepackage{amsmath}
\usepackage{mathrsfs}
\usepackage{hyperref}
\usepackage{multirow}
\usepackage{subfigure}
\def\be{\begin{equation}}
\def\ee{\end{equation}}
\def\ba{\begin{eqnarray}}
\def\ea{\end{eqnarray}}
\def\nl{\nonumber\\}
\def\aa{a}

\def\ZZ{\mathcal{Z}}

\def\Li{\textrm{Li}}
\def\l{\langle}
\def\r{\rangle}
\def\SS{\mathcal{S}}

\def\Roneone{R${}^{1,1}$~}
\def\BB{\mathcal{B}}

\title{Three-loop octagons and $n$-gons in maximally supersymmetric Yang-Mills theory}
\author{Simon Caron-Huot${}^{a,b}$ and Song He${}^{c,b}$}
\affiliation[a]{Niels Bohr International Academy and Discovery Center, Blegdamsvej 17, Copenhagen 2100, Denmark} \affiliation[b]{School of Natural Sciences,
Institute for Advanced Study, Princeton, NJ 08540, USA} \affiliation[c]{Perimeter Institute for Theoretical Physics, Waterloo, ON N2L 2Y5,
Canada} \emailAdd{schuot@nbi.dk, she@perimeterinstitute.ca}\abstract{We study the S-matrix of planar $\mathcal{N}=4$ supersymmetric Yang-Mills
theory when external momenta are restricted to a two-dimensional subspace of Minkowski space. We find significant simplifications and new,
interesting structures for tree and loop amplitudes in two-dimensional kinematics; in particular, the higher-point amplitudes we consider can be obtained from those with lowest-points by a collinear uplifting. Based on a compact formula for one-loop N${}^2$MHV amplitudes, we use an equation proposed
previously to compute, for the first time, the complete two-loop NMHV and three-loop MHV octagons, which we conjecture to uplift to give the
full $n$-point amplitudes up to simpler logarithmic terms or dilogarithmic terms.}

\begin{document}
\maketitle
\section{Introduction}

Recently, tremendous progress has been made to understand the rich structure of scattering amplitudes in gauge theories, especially
in planar $\mathcal{N}=4$ supersymmetric Yang-Mills theory (SYM). For instance, the on-shell diagram method has drastically improved our
knowledge of mathematical structures underlying scattering amplitudes, and the all-loop integrand for planar $\mathcal{N}=4$ SYM has been
obtained (see~\cite{Grassmannian, ArkaniHamed:2010kv,ArkaniHamed:2010gh,ArkaniHamed:2012nw} and references therein). There has been a lot of significant progress in computing S-matrix itself analytically to relatively high loop-orders, see for example ~\cite{DelDuca:2010zg,Goncharov:2010jf,arXiv:1105.5606,arXiv:1006.4127,khoze,Alday:2010jz,Dixon:2011pw,Heslop:2011hv,Dixon:2011nj,CaronHuot:2011kk}
Very recently, based on the integrability of the theory~\cite{minahanzarembo,beisertstaudacher}~\cite{Beisert:2010jr}, it was proposed~\cite{Basso:2013vsa} that
computations for the operator product expansion (OPE) of amplitudes/Wilson loops~\cite{Alday:2010ku} can be done at finite couplings!

\begin{figure}\centering
\includegraphics[height=2.5cm]{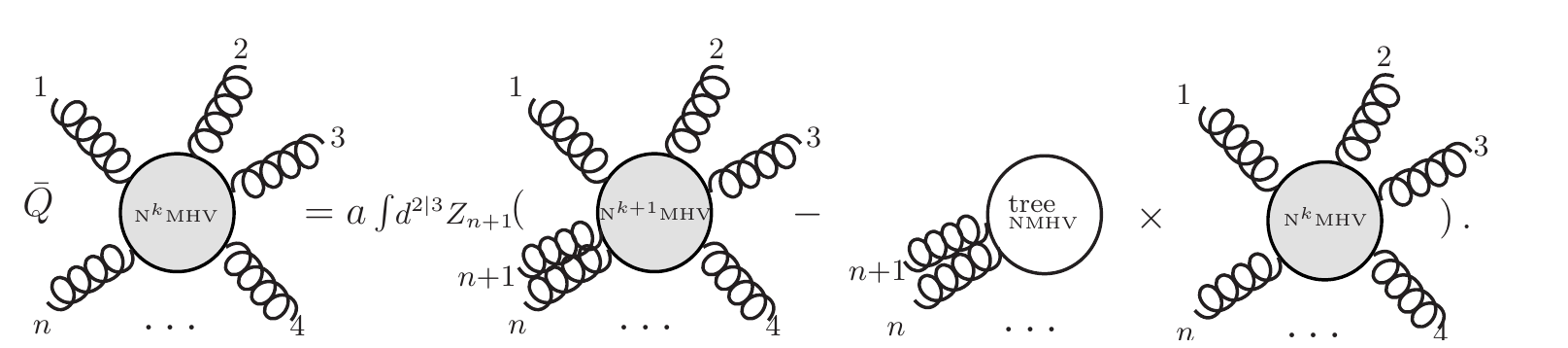}
\caption{All-loop $\bar Q$ equation for planar $\mathcal{N}=4$ S-matrix.} \label{fig:equation}
\end{figure}

It is well known that in planar $\mathcal{N}=4$ SYM, scattering amplitudes are dual to null polygonal Wilson loops in a dual spacetime
~\cite{aldaymalda,Drummond:2007aua,WL1,WLDrummond,Bern:2008ap,WLDrummond2,maldacenaberkovits,masonskinner,arXiv:1010.1167}. They enjoy both superconformal symmetries in the original and dual spacetime (known as dual conformal symmetry~\cite{magic,aldaymalda}), which close into the
infinite-dimensional Yangian symmetry~\cite{Yangian}. In ref.~\cite{CaronHuot:2011kk}, \emph{exact} all-loop equations obeyed by the S-matrix
were derived, by determining the quantum corrections to the generators of Yangian symmetry acting on the (BDS-renormalized) S-matrix. The
equations consist of a $\bar Q$ equation for the dual supersymmetry~\cite{sokatchevDCI},  and its
parity-conjugate~\cite{CaronHuot:2011kk}; see Fig.~\ref{fig:equation}.

The equations express the derivatives of amplitudes at a given loop order in terms of integrals of lower-loop amplitudes with more legs, and at
least perturbatively, they can be solved to uniquely determine the S-matrix. Using this technique, MHV and NMHV at two loops at six-point (hexagons), were derived reproducing results in~\cite{Goncharov:2010jf} and~\cite{Dixon:2011nj}. The complete symbol~\footnote{The symbol technique is a powerful tool to deal with iterated integrals,
see~\cite{Goncharov:2010jf}.} of two-loop NMHV heptagon and three-loop MHV hexagon was determined for the first time. However, the formula for multi-loop multi-particle amplitudes are fairly complicated, e.g. two-loop MHV becomes quite involved as the number of points increases, and it seems beyond our present reach to compute and understand, in an analytic form, the non-trivial scattering process at higher points or higher loops.

This is one of the motivations to consider restricted kinematic configurations in which scattering amplitudes/Wilson loops simplify.
Following~\cite{Alday:2009ga,Alday:2009yn} at strong coupling, and~\cite{arXiv:1006.4127,khoze} at weak coupling, we wish to consider
configurations of external momenta/edges of the Wilson loop which lie in 1+1 dimensional subspace of the Minkowski space. It is well known that
amplitudes/Wilson loops simplify a lot in such \Roneone kinematics: the $n$-gon Wilson loops (dual to $n$-point, MHV amplitudes) have
been computed at one and two loops~\cite{khoze,arXiv:1010.5009}, and the computation is expected to be much simpler also at strong coupling ~\cite{Alday:2009ga,Alday:2009yn,Alday:2010vh}.

We strongly believe that we have only began to see a rich venue of analytic formula of scattering amplitudes in \Roneone kinematics. As we will demonstrate in this paper, not only we obtain very compact formula with huge simplifications when we go to higher loops or go deeper into non-MHV
sectors, but we also see non-trivial structures, invisible in previous calculations for \Roneone amplitudes, which begin to emerge there.
Therefore, although simpler the results still exhibit considerable complexity, making the \Roneone kinematics an ideal laboratory. In
particular, as we will demonstrate in this paper, it appears that the essential subtleties involving fermionic modes left
unresolved in the recent OPE proposal of~\cite{Basso:2013vsa} are fully present already in \Roneone kinematics.

The all-loop equations of~\cite{CaronHuot:2011kk} also simplify in two dimensions, and in particular the $\bar Q$ equation is closed within
the \Roneone kinematics. In this previous work we initiated the study of non-MHV amplitudes (supersymmetrically) reduced to two dimensions, and
obtained an analytic expression for $n$-point one-loop NMHV amplitudes. In this paper, combining the $\bar Q$ equation and an uplifting formalism proposed \cite{Goddard:2012cx}, we will push the study to the next loop order by presenting, for the first time, compact analytic formula for one-loop N${}^2$MHV, two-loop NMHV and three-loop MHV amplitudes, in a manifestly dual superconformal form. In particular,  the new results for the octagons at one, two and three loops can be found in eqs.~(\ref{8ptN2MHV}), (\ref{mainresult2loop}) and (\ref{3loopoctagon})-(\ref{mainresult3loop}) respectively.

Two new observations about the results are worth emphasizing here. First, unlike the $k{+}\ell=2$ case, general two-dimensional amplitudes are
``non-factorizable", \emph{i.e.} they must contain functions that mix distinct conformal cross-ratios. The one-loop N${}^2$MHV octagon is the
first example in \Roneone kinematics where non-trivial factors, such as the difference of two cross-ratios, appear in the denominator. The
mixing of cross-ratios also emerges at higher loops, and we believe this structure to be universally present in multi-loop amplitudes.

Furthermore, we will find that generally \Roneone higher-point amplitudes can be uplifted from lower-point building blocks, which e.g. can be identified already in the octagon. In a precise sense, the ``most complicated'' part of the $n$-point amplitude is obtained by uplifting the octagon.
This method of uplifting was first proposed in~\cite{Goddard:2012cx} based on the structure of the two-loop MHV, and
is designed to trivialize the collinear-soft limits of amplitudes in \Roneone\!

The paper is organized as follows. We introduce our notations and basic ingredients for amplitudes in \Roneone kinematics, in
section~\ref{preliminaries}. In section~\ref{1loopNMHV} we review the collinear-soft uplifting formalism, and use it to present a manifestly
conformal invariant formula for all one-loop NMHV amplitude. In section~\ref{N2MHV}, we derive the one-loop N${}^2$MHV octagon by reducing the
four-dimensional expression, and uplift it to higher points. Based on the result, in section~\ref{2loopNMHV}, we use
$\bar Q$ equation to obtain two-loop NMHV amplitudes: we first discuss in details how the equation works (\ref{LHS} and \ref{RHS}), then we use
the technique described in section~\ref{integratesymbol} to obtain the octagon (\ref{8ptNMHV}), and proceed to the uplifting (\ref{10ptNMHV}).
Finally we upgrade the two-loop NMHV using $\bar Q$ equation again and present the three-loop MHV octagon in section~\ref{3loop}; we
also discuss numerical comparisons and the OPE expansion of the results, and include some useful materials in the appendix.

\section{Preliminaries}\label{preliminaries}

\subsection*{\Roneone kinematics}
In two dimensions, without loss of generality, we can always take the number of particles, $n$, to be even. The Wilson loop contour, which is a polygon with $n$ light-like edges in the dual space, must take a zigzag shape~\cite{Alday:2009ga,Alday:2009yn}. In the light-cone coordinate,
$x=(x^+, x^-)$ with $x^{\pm}=x^0\pm x^1$, the $n$ cusps can be parametrized as follows : \be
x_{2i{-}1}=(X_{2i{-}1},X_{2i{-}2}),\qquad x_{2i}=(X_{2i{-}1},X_{2i}),\ee for $i=1,...,\frac n 2$, where odd and even labels of cusps are
distinguished, see Fig.~\ref{fig:BCFW}. Given that $x_{i{+}1}-x_i=p_i$, any external momentum only depends the difference of two $X$'s of the same parity,
$p_{2i{-}1}=(0,X_{2i}-X_{2i{-}2})$, $p_{2i}=(X_{2i{+}1}-X_{2i{-}1},0)$. This reduces the conformal group SU(2,2) to SL(2)$\times$SL(2).
Actually, we are interested in super-amplitudes, and it is very natural to consider a supersymmetric reduction from SU(2,2$\mid$4) to
SL(2$\mid$2)$\times$SL(2$\mid$2)~\cite{CaronHuot:2011kk}.

Scattering amplitudes/Wilson loops in planar $\mathcal{N}=4$ SYM are most conveniently present using the so-called momentum twistors~\cite{hodges}, for which we recall the definition:
\be
\ZZ=(Z^a|\eta^A):=(\Lambda^{\alpha},x^{\alpha \dot\alpha}
\Lambda_{\alpha}|\theta^{\alpha A} \Lambda_{\alpha})~.\ee
In \Roneone kinematics, as the symmetry reduction implies, two components of the bosonic twistor (and two of the fermionic part) vanish, \footnote{Here and in the rest of the paper we use $\lambda$ and $\tilde{\lambda}$ to denote the non-vanishing components of odd and even momentum twistors, respectively, which should not be confused with the spinors in four dimensions, $\Lambda$ and $\tilde{\Lambda}$.}
\be
\ZZ_{2i{-}1}=(\lambda^1_{2i{-}1},0,\lambda^2_{2i{-}1},0|\chi^1_{2i{-}1},0,\chi^2_{2i{-}1},0), \qquad
\ZZ_{2i}=(0,\tilde{\lambda}^1_{2i},0,\tilde{\lambda}^2_{2i}, 0|0,\tilde{\chi}^1_{2i},0,\tilde{\chi}^2_{2i})\,.\label{mom-twistor}
\ee
This will be our definition of \Roneone kinematics.

It is apparent that one can write SL(2)-invariants for either odd or even particles. The invariant two-bracket for twistors in the odd sector is
defined as $\l i\,j\r:=\epsilon_{\alpha,\beta}\lambda^\alpha_i\lambda^\beta_j$ when $i$ and $j$ are odd, and similarly in the even sector
$[i\,j]:=\epsilon_{\alpha,\beta}\tilde{\lambda}^\alpha_i\tilde{\lambda}^\beta_j$. These are in fact one-dimensional distance in $p^-$ ($p^+$)
direction for the odd (even) sector, $\l i\,j\r=X_i-X_j$.  Any expression in terms of SL(2,2)-invariant four-brackets in four-dimensions,
$\l i\,j\,k\,l\r:=\epsilon_{a,b,c,d} Z^a_i Z^b_j Z^c_k Z^d_l$, can be easily reduced to \Roneone kinematics: the only non-vanishing four-bracket
involves two odd labels, e.g. $i,k$, and two even ones, $j,l$, \be \l i\,j\,k\,l \r =\l i\,k\r [j\,l]. \ee

\subsection*{Tree amplitudes}
The MHV-stripped tree-level amplitudes are Yangian invariant. The MHV amplitude is, by definition, unity: $R^\textrm{tree}_{n,0}:=1$. For $k>0$,
it is useful to write basic invariants (called R-invariants) under the odd-sector SL(2$\mid$2),  \be (i\,j\,k) := \frac{\delta^{0\|2}\big(\l i\,j\r
\chi_{k} +\l j\,k\r \chi_{i}+\l k\,i\r \chi_{j}\big)}{\l i\,j\r\l j\,k\r\l
 k\,i\r},\label{tripleRinv}
\ee and similarly R-invariants in the even sector with $\l\,\r \leftrightarrow [\,]$. We find it useful to encode the odd/even information into
the notation: we write $(i)$ with a parenthesis when $i$
is odd, and denote the same object as $[i]$ when $i$ is even. These invariants satisfy $(i\,j\,k)+(j\,k\,l)+(k\,l\,i)+(l\,i\,j)=0$. The simplest example is the 6-point NMHV tree, 
\be R^{\textrm{tree}}_{6,1}=-(1\,3\,5)[2\,4\,6]. \ee

Here we present the two-dimensional BCFW recursion relations, which can be used to generate all $n$-point, N${}^{k{-}2}$MHV amplitudes in
\Roneone kinematics. The derivation of the formula and more details of tree amplitudes in two dimensions can be found in
Appendix~\ref{app:tree}. \be
\begin{array}{l}\displaystyle
 R^\textrm{tree}_{n,k}(1,\ldots,n) = R^\textrm{tree}_{n{-}2,k}(1,\ldots,n-2)
\\\displaystyle
 +\sum_{\substack{i=3\\i~\textrm{odd}}}^{n{-}3}\sum_{k_L=0}^{k-1}
   (n{-}1\,1\,i)\big( [n{-}2\,n\,i{+}1]-[n{-}2\,n\,i{-}1]\big)R^\textrm{tree}_{i{+}1,k_L}(1,2,\ldots,i,n) R^\textrm{tree}_{n{-}i{-}1,k_R}(i,\ldots,n{-}2)
\,,
\end{array}\label{BCFW}
\ee where $k_R=k{-}1{-}k_L$, and it is understood that $R^\textrm{tree}_{2,0}=1$ for the term with $n{-}i{-}1=2$. Notice that, in
contradistinction with the four-dimensional case, the arguments of the amplitude remain \emph{unshifted}. This is a special feature of \Roneone
kinematics and originates from the fact that all products of two-brackets which arise can be simplified via Schouten identities.

A particularly nice example is the 8-point N${}^2$MHV tree,\be
 R^\textrm{tree}_{8,2} = (1\,3\,5)(5\,7\,1)[4\,6\,8][8\,2\,4] = (1\,3\,5\,7)[2\,4\,6\,8]\,,
\ee
where the first equality follows from eq.~(\ref{BCFW}); in the second equality, we have defined higher-order R-invariants by $(i\,j\,k\,l)=-(i\,j\,k)(k\,l\,i)$, or explicitly,
\be
 (i\,j\,k\,l):=\frac{\delta^{0\|4}\big(\l i\,j\r \chi_{k}\chi_l +\l j\,k\r \chi_{l}\chi_i+\l k\,l\r \chi_{i}\chi_j+\l l\,i\r\chi_j\chi_k\big)}{\l i\,j\r\l j\,k\r\l
 k\,l\r\l l\,i\r}\,.
\label{quardRinv}\ee This rewriting makes the formula manifestly cyclic invariant.

\subsection*{Loop amplitudes}
At loop level, we will exploit the dual conformal invariance to write (BDS-subtracted) amplitudes as functions of conformal cross-ratios. Let us first define the most general cross-ratios in \Roneone kinematics, \be v_{i\,j\,k\,l}:=\frac{\l i\,j\r \l k\,l\r}{\l i\,l\r \l j\,k\r}, \ee for $i,j,k,l$ being odd, and $\l\,\r\leftrightarrow [\,]$ for all labels even. Note that one can not construct two-dimensional cross-ratios with mixed parity.

Any cross-ratio in four-dimensional kinematics factorizes, \emph{i.e.} it reduces to the product of an odd and an even cross-ratios. For
instance, when $i,j,k,l$ are of the same parity, \be \frac{x^2_{i,j}x^2_{k,l}}{x^2_{i,l}x^2_{j,k}} \to v_{i{-}1,j{-}1,k{-}1,l{-}1}v_{i,j,k,l}
.\ee

It is important to understand the Euclidean region of the kinematics, where the functions are real. In the Euclidean region, all non-adjacent
cusps are space-like separated, and all odd/even edges move in the same direction. In terms of the light-cone coordinate, all cusps of the same
parity must be cyclically ordered, \emph{i.e.} $X_i < X_j$ for $i<j$ ($i,j$ of the same parity) and when we go from $x_n$ to $x_1$ along
edge $n$, $X_{n{-}1}$ must go along the same direction, ``wrap around infinity", to reach $X_1$, see \cite{arXiv:1010.5009}. In terms of
cross-ratios, the condition translates to $v_{i,j,k,l}$ being strictly positive when $i,j,k,l$ are cyclically ordered; thus the Euclidean region
is given by $\{v_{i,j,k,l}>0, \forall~i{<}j{<}k{<}l{<}i\}$.

One-loop MHV amplitudes have been known explicitly for a while~\cite{bddk2}. They simplify even further when restricted to two-dimensional kinematics, all the dilogarithms cancel~\cite{khoze}, and one is left with
\be
 \frac{A_{n,0}^\textrm{1-loop}}{A_{n,0}^\textrm{tree}} =
 \sum_{i~\textrm{odd}}\sum_{\substack{j=i{+}3\\j~\textrm{even}}}^{i{-}3} \log \frac{\l i\,j{-}1\r}{\l i\,j{+}1 \r} \log\frac{[i{-}1\,j]}{[i{+}1\,j]}
 -\frac12 \sum_{i=1}^n \log^2\frac{x_{i,i{+}2}^2}{\mu^2_\textrm{IR}}\label{1loopMHV}
 \,,
\ee
where, in the Euclidean region, the arguments of the logarithms should be taken in absolute value.
In the rest of this paper we will deal with the remainder function and all the arguments of logarithms will be cross-ratios
(which are positive in the Euclidean region), so this qualification will not be needed.

The one-loop MHV amplitude enters the definition of the so-called BDS-subtracted amplitude or remainder function. The exponentiated Bern-Dixon-Smirnov Ansatz was proposed in~\cite{Bern:2005iz}, based on the iterative relation~\cite{Anastasiou:2003kj}, and it captures the infrared and collinear behavior of general amplitudes to all loops. By subtracting the BDS expression, the answer becomes infrared finite, dual conformal invariant and has simple collinear limits;  in \Roneone kinematics we explicitly define~\cite{CaronHuot:2011kk} the $n$-point, N${}^{k{-}2}$MHV, BDS-subtracted amplitude by
\be
 A_{n,k} := e^{\Gamma_\textrm{cusp} \frac{A_{n,0}^\textrm{1-loop}}{A_{n,0}^\textrm{tree}} + c_1\sum_i \log \frac{\mu^2_\textrm{IR}}{x^2_{i,i{+}2}}+c_2 +(c_3+c_3') (n-4) +c_4' k} R_{n,k}\,,\label{BDS}
\ee
where we define the coupling $g^2\equiv\frac{g^2_\textrm{YM}N_c}{16\pi^2}$, and $\aa=\Gamma_\textrm{cusp}=g^2-\frac{\pi^2}3 g^4+\ldots$ is the cusp anomalous dimension, which will be used as our loop expansion parameter, \be R_{n,k}=\sum_{\ell=0}^{\infty}\aa^{\ell}~R^{(\ell)}_{n,k}.\ee $c_1,\ldots,c_4'$ are coupling-constant dependent (and in some case, scheme dependent) coefficients. While $c_1,c_2,c_3$ are related to the usual coefficients entering the BDS Ansatz \cite{Bern:2005iz}, we stress that coefficients $c_3', c_4'$ are specific to \Roneone kinematics.
They define a rescaling of the BDS-subtracted amplitude, such that it has simpler \Roneone collinear-soft limits (see below). The coefficients can be uniquely determined by the requirements that \be R_{6,0}=1 \quad\mbox{and}\quad R_{6,1}=R^{\textrm{tree}}_{6,1}\,,\ee to all loop orders in \Roneone kinematics, and explicitly we know~\cite{arXiv:1006.4127,CaronHuot:2011kk}
\be c_3'=-\aa^2\frac{\pi^4}{18} +\mathcal{O}(\aa^3), \quad c_4'=-\aa\frac{\pi^2} 3 +\aa^2 \frac{7 \pi^4}{30}+\mathcal{O}(\aa^3).\ee

\section{An invitation: all one-loop NMHV amplitudes from the octagon}\label{1loopNMHV}

We will now present the one-loop correction to the $n$-point NMHV (BDS-subtracted) amplitude in \Roneone kinematics. Although they have been computed previously in ref.~\cite{CaronHuot:2011kk}, here we will rewrite the expression in a much nicer form, which manifests the dual (super)conformal symmetry and the dihedral symmetry of the problem. This rewriting will be based on the powerful collinear-soft uplifting formalism of ref.~\cite{Goddard:2012cx}, and we will find that all $n$-point one-loop NMHV amplitudes are obtained by uplifting the 8-point one.


The one-loop NMHV result of \cite{CaronHuot:2011kk}, in the original notation, where $(i)$ means (2i{-}1), $[i]$ means $[2i]$ and odd/even cross-ratios are denoted as $u^{\pm}$,  was written as:
\ba
 \tilde R_{2n,1}^\textrm{(1)} &=& \!\!\!\!\!\!\!\sum_{i<j<k<l<i} (j\,k\,l)([j\,l{-}1\,l]-[j{-}1\,l{-}1\,l])\log \l ik\r \log u_{i{-}1,k{-}1}^-
-\!\!\!\! \sum_{i<j<k<i} \log \l i k\r (i\,j\,k) \nl
 &&  \hspace{-0.5cm}\times \left( \log u^-_{i,k{-}1,k,j}( [i\,j{-}1\,j]-[i{-}1\,j{-}1\,j]) + \log u^-_{k{-}1,i{-}1,i,j}([k\,j{-}1\,j]-[k{-}1\,j{-}1\,j]) \right.
 \nl && + \left.\left( \log u^-_{i,j,j{-}1,k} [i\,j{-}1\,k] - (i\leftrightarrow i{-}1)-(j\leftrightarrow j{-}1)\right) \right)\,.\label{NMHV1loop}
\ea
The compactness of this expression is deceiving, as it artificially contains ill-defined $\log[i\,i]$ terms which cancel out in the sum, and are to be dropped.
The result is a conformally invariant function, although no attempt was made in~\cite{CaronHuot:2011kk} to express it directly in terms of cross-ratios. We will now do so for the octagon, and then uplift the result to $n$-point.

\subsection{The 8-point NMHV amplitude}\label{octagonNMHV}

It is useful to discuss the general structure of the NMHV octagon to all loop orders. Due to unbroken $Q$ (dual) superconformal symmetry, at any loop order the helicity dependence of NMHV amplitudes is encoded in the R-invariants defined in eq.~(\ref{tripleRinv}). At 8-points it is convenient to further label them as $(i) := (i\,i{+}2\,i{+}4)$ and $[j]:=[j\,j{+}2\,j{+}4]$, for $i$ odd and $j$ even. They obey the four-term identities $(1)-(3)+(5)-(7)=[2]-[4]+[6]-[8]=0$. It follows that there are 9 linearly independent invariants for NMHV 8 point, obtained by taking a product of an odd invariant with an even invariant.

To choose a basis for these invariants, it is helpful to note that $9=8+1$. This trivial fact suggests to pick a basis consisting of one cyclic class of length 8 plus one cyclic-invariant combination, which is naturally chosen to be the tree amplitude given in eq.~(\ref{nmhvtree2d})
\be
 R_{8,1}^\textrm{tree} = (3)[6]-(3)[4]-(1)[8]-(5)[6]\,.
\ee
Furthermore, by dual conformal symmetry, the (bosonic) functions multiplying each invariant depends only on the two independent cross-ratios
\be
 v_1 = \frac{\l 1\,3\r\l5\,7\r}{\l 1\,7\r\l 3\,5\r}, \qquad v_2 = \frac{[2\,4][6\,8]}{[2\,8][4\,6]}\,.
\ee
We remark that any function of these cross-ratios is automatically invariant under cyclic rotations by 4.

It follows that the 8-point NMHV amplitude can be expressed, to any loop order, as
\be
 R_{8,1} = \big( \big([2](3) +[6](7)\big)f_{8,1}^1(v_1,v_2) + \textrm{3 cyclic} \big) + R_{8,1}^\textrm{tree} f_{8,1}^2(v_1,v_2). \label{R81}
\ee
Expanding out the formula explicitly,
\ba
 R_{8,1} &=& \big( [2](3)+ [6](7) \big) f^1_{8,1}(v_1,v_2) +\big( (3)[4] + (7)[8] \big) f^1_{8,1}(v_2,\frac 1 {v_1})\nl
 &&\big( [4](5)+[8](1)\big) f^1_{8,1}(\frac 1 {v_1},\frac 1 {v_2}) +\big((5)[6]+(1)[2] \big) f^1_{8,1}(\frac 1 {v_2},v_1)\nl
  && + \big( (3)[6]-(3)[4]-(1)[8]-(5)[6] \big) f^2_{8,1}(v_1,v_2). \label{eightpointNMHV}
\ea
Note that various arguments of $f^1$  simply follow from the transformation of the cross-ratios under cyclic rotations. Dihedral symmetry implies that $f^1_{8,1}(v_1,v_2)=f^1_{8,1}(v_2,v_1)$.
Furthermore, the coefficient of the tree amplitude enjoys the full dihedral symmetry, $f^2_{8,1}(v_1,v_2)=f^2_{8,1}(v_2,v_1)=f^2_{8,1}(\frac 1{v_1},v_2)$.

At tree level we have by definition $f^{2(0)}_{8,1}=1$ and $f^{1(0)}_{8,1}=0$.
By evaluating the general one-loop formula (\ref{NMHV1loop}) on various components, one can extract the values of the $f_{8,1}^i$ at one-loop.
For example, the component $\chi_1\chi_3\chi_4\chi_6$ gives $f^2_{8,1}$ up to a normalization factor etc.
This way we obtain
\begin{align}
 f^{1(1)}_{8,1}(v_1,v_2) &= \log (1+v_1)\log(1+v_2)\label{oneloop81a} \\ \label{oneloop81b}
 f^{2(1)}_{8,1}(v_1,v_2) &= \big(\log (1+v_1)+\log(1+\frac 1 {v_1})\big) \big(\log (1+v_2)+\log(1+\frac 1 {v_2})\big).
\end{align}
This describes completely the one-loop NMHV octagon.

Let us comment on the collinear limits of the 8-point amplitude. In \Roneone kinematics, the natural collinear limit is the collinear-soft limit: when taking $p_{i{-}1}$ and $p_{i{+}1}$ to be collinear, the momentum of the particle in between becomes soft, $p_i\rightarrow 0$. The behavior of general amplitudes in this limit is well understood~\cite{khoze,DelDuca:2010zg} and particularly simple due to the definition (\ref{BDS}):
\be
\lim_{ z_n\to z_{n{-}1}} R_{n,k}(1,...,n) =R_{n{-}2,k}(1,...,n{-}2),
\ee
and a similar $k$-decreasing limit.
For definiteness, we can consider here the $k$-preserving limit in which $z_7\to z_5$ so that $v_1\to 0$, while the even twistors are held fixed.
In this limit $(3\,5\,7)$ and $(5\,7\,1)$ disappears while $(7\,1\,3)\to(1\,3\,5)$.  Collecting the terms proportional to $(1\,3\,5)$, one finds a combination of $[2\,4\,6]$, $[4\,6\,8]$ or $[8\,2\,4]$. These three invariants are linearly independent so each of their coefficient must vanish separately; this imposes
\ba
  0 &=&\lim_{v_1\to 0} f^1_{8,1}(v_1,v_2),  \label{coll81a}\\
 1 &=& \lim_{v_1\to \infty}(f^2_{8,1}(v_1,v_2)-f^1_{8,1}(v_1,v_2)-f^1_{8,1}(v_1,\frac 1 {v_2})).\label{coll81b}
\ea
Other limits such as the $k$-decreasing ones lead to identical constraints.
The amplitude (\ref{oneloop81a})-(\ref{oneloop81b}), up to an overall proportionality constant, is the only amplitude of transcendental degree two that is consistent with these limits, with dihedral symmetry, and whose only possible branch points in the physical region are at $v_1=-1,0,\infty$ and $v_2=-1,0,\infty$ (as required by unitarity).
\subsection{Collinear-soft uplifting formalism}\label{uplift}

In~\cite{Goddard:2012cx}, it was proposed that $n$-point amplitudes in \Roneone kinematics can be obtained as collinear-soft uplifting of low-point objects, which are to be viewed as a function of off-shell points rather than polygons.
In general, the uplifting begins with a function $S_8$ which depends on four (off-shell) points, but
may involve higher partial amplitudes like $S_{10},S_{12}$ etc., depending on the loop order:
\def\ll{\ell}
\be R_n=\sum_{1\leq i\lhd j\lhd k\lhd \ll\leq n} (-)^{i{+}j{+}k{+}\ll} S_8(x_i,x_j,x_k,x_\ll)+\mbox{contributions from $S_{10}, S_{12}$ etc}.\label{generaluplift}\ee
The symbol $i\lhd j$ means that the indices should be separated by at least 2, $j\geq i+2$ (although not made explicit in the notation, it is understood that $\ll$ and $i$ must be separated also, e.g. when $i=1$ we must have $\ll\leq n-1$).

The basic objects are largely restricted by physical constraints. For instance, $S_8$ should be a cyclically invariant function of 4 cusps. Concretely, each cusp $x_i:= (Z_{i-1},Z_i)$ is associated to one odd and one even twistor and so we will write
\be
 S_8(x_i,x_j,x_k,x_\ll) := S_8(i_\textrm{odd}, i_\textrm{even},j_\textrm{odd}, j_\textrm{even},k_\textrm{odd}, k_\textrm{even},\ll_\textrm{odd}, \ll_\textrm{even})\,.
\ee
Then $S_8$ must be invariant under cyclic rotations of the twistors by \emph{two}.

Furthermore, we require invariance under interchange of odd and even labels,\\
$S_8(1,2,3,4,5,6,7,8)=S_8(2,1,4,3,6,5,8,7)$.  These two properties ensure that the $S_8$ contribution in (\ref{generaluplift}) is cyclically invariant for all $n$.  As far as possible, we would also like $S_8$ to vanish in collinear limits so as to make the collinear limits as simple as possible:
\be
 S_8(x_i,x_j,x_k,x_{\ll})\to 0, \qquad \mbox{if $x_k,x_{\ll}$ are light-like separated}\,.
\ee
However we will see that this may not always be achievable, e.g. for two-loop NMHV and three-loop MHV. In such cases, as we will show below, a natural modification to eq.~(\ref{generaluplift}) is needed: we simply include the non-vanishing collinear limits, e.g. $S(x_i, x_j, x_k, x_{k{+}1})$, by relaxing the off-shell condition, e.g. $k\lhd \ll$, in the sum!

Finally, (\ref{generaluplift}) has to give the correct 8-point amplitude
\be
 R_{8} = S_{8}(x_2,x_4,x_6,x_8)+S_{8}(x_1,x_3,x_5,x_7) \,, \label{matching8point}
\ee
which is equivalent to
\be
 R_8 = S_8(1,2,3,4,5,6,7,8) + S_8(8,1,2,3,4,5,6,7)\,.
\ee

One of the motivations for this formalism in ref.~\cite{Goddard:2012cx} was the extremely simple expression for the two-loop MHV remainder function
in \Roneone kinematics: the $n$ point MHV amplitude is described simply by using
\be
 S_{8,0}^{(2)} = -\log (1+v_1)\log (1+\frac 1{v_1}) \log (1+v_2)\log (1+\frac 1{v_2})\label{uplift8}
\ee
in eq.~(\ref{generaluplift}), and setting $S_{10}=S_{12}=\ldots=0$.
It is remarkable that $S^{(2)}_{8,0}$ takes the simplest possible form, namely the only combination of logarithms with physically allowed arguments
and which vanishes in collinear limits.

In ref.~\cite{Goddard:2012cx} it was conjectured that a similar uplifting formalism should exist for more general amplitudes. In the next subsection
we will confirm this conjecture in the case of one-loop NMHV amplitudes. 

\subsection{The $n$-point one-loop NMHV amplitude from uplifting}\label{upliftNMHV}

Following (\ref{matching8point}), $S_{8,1}^{(1)}$ is obtained by ``splitting'' $R_{8,1}^{(1)}$ into two parts.
It is important to note that one cannot simply take $S_{8,1}^{(1)}=\frac12 R_{8,1}^{(1)}$: this has the wrong symmetries.
The behavior of the amplitude (\ref{eightpointNMHV}) in the collinear-soft limits gives us a hint as to how to split it,
as it is easy to see that there is no interplay between the different logarithms in the limit.
Thus a natural Ansatz for $S_{8,1}^{(1)}(x_2,x_4,x_6,x_8)=S_{8,1}^{(1)}(1,2,3,4,5,6,7,8)$ is to collect half the logarithms in $R_{8,1}^{(1)}$:
\ba
 S_{8,1}^{(1)}&=& \phantom{+}
 \big( (1\,3\,5)-(7\,1\,3) \big)\big( [8\,2\,4]-[2\,4\,6]\big) \log (1+\frac 1{v_1})\log (1+\frac 1 {v_2})
\nl &&+
\big( (3\,5\,7)-(1\,3\,5) \big)\big( [2\,4\,6]-[4\,6\,8]\big) \log (1+v_{1})\log (1+v_{2})\,.
\label{S81}\ea

Just like $R^{(2)}_{8,0}$, these building blocks can be identified as the most general functions which vanish in collinear limits.
For each term and in each limit, for instance the supersymmetric limits $z_3\to z_1$ or $z_5\to z_3$, one finds that either the R-invariants or the logarithm vanish.  Furthermore, $S_{8,1}^{(1)}$ is invariant under cyclic rotations by 2 and under interchange of odd and even labels,
as required. Finally, $S_{8,1}^{(1)}(1,2,3,4,5,6,7,8)+S_{8,1}^{(1)}(8,1,2,3,4,5,6,7)$ reproduces the 8-point amplitude (\ref{eightpointNMHV}), as is easily verified.
Thus this provides an essentially unique Ansatz for $S_{8,1}^{(1)}$.

Amazingly, plugging this $S_{8,1}^{(1)}$ into (\ref{generaluplift}), we find that it reproduces precisely
all $n$-point amplitudes given by (\ref{NMHV1loop})! Thus eq.~(\ref{S81}) together with eq.~(\ref{generaluplift}) (with $S_{10}=S_{12}=\ldots=0$) give an explicit and compact description of all one-loop $n$-point  NMHV amplitudes.

\section{One-loop N${}^2$MHV amplitudes and uplifting}\label{N2MHV}

In this section we will give new results on one-loop N${}^2$MHV amplitudes. We study the reduction of four-dimensional octagon, and obtain a
very compact expression for the octagon in two-dimensional kinematics. We further show that there is again a powerful uplifting formalism.
N${}^2$MHV octagons and decagons have been studied using OPE in~\cite{Sever:2012qp}, where our results below have been checked.

\subsection{Dual-conformal box-expansion and one-loop N${}^2$MHV octagon}\label{8ptNNMHV}

Here we recall the main result of~\cite{Bourjaily:2013mma}, that in the box expansion for the one-loop ratio function the box integrals can be replaced by the
following, simpler and more uniform, expressions: \ba
 \tilde F^{3m}_{\{i{-}1\,i\}\,j\,k} &=& \Li_2(1-u_{i\,j\,k\,i{-}1}) +\frac12\log u_{i\,j\,k\,i{-}1} \log v^{(3)} \nonumber\\
 \tilde F^{2me}_{\{i{-}1\,i\}\,j{-}1\,j} &=& \Li_2(1-u_{i\,j{-}1\,j\,i{-}1}) +\frac12\log u_{i\,j{-}1\,j\,i{-}1} \log v^{(2e)} \nonumber\\
 \tilde F^{2mh}_{\{i{-}1\,i\}\,i{+}1\,j} &=& \Li_2(1) +\frac12\log u_{i\,i{+}2\,j\,i{-}1} \log u_{i\,i{-}2\,j\,i{+}1} \nonumber\\
 \tilde F^{1m}_{\{i{-}1\,i\}\,i{+}1\,i{+}2} &=& \Li_2(1) \label{DCIboxes}
\ea
where $u_{i\,j\,k\,l}$ are the conventional cross-ratios in four dimensions:
\def\s{x^2}
$u_{i\,j\,k\,\ll}:=\frac{\s_{i\,j}\s_{k\,\ll}}{\s_{i\,k}\s_{j\,\ll}}$,
and
\ba
\nonumber
 v^{(3)} &=&
 \frac{\s_{j\,k}\s_{i{-}2\,i}\s_{i{-}1\,i{+}1}}{\s_{i{-}1\,j}\s_{i\,k}\s_{i{-}2\,i{+}1}}\,,
 \quad
 v^{(2e)}=\frac{\s_{i{-}2\,i}\s_{j{-}2\,j}\s_{i{-}1\,i{+}1}\s_{j{-}1\,j{+}1}}{\s_{i{-}1\,j{-}1}\s_{i\,j}\s_{i{-}2\,i{+}1}\s_{j{-}2\,j{+}1}}\,,
\ea
The four-mass-box, which first appear in N${}^2$MHV octagon, is given as
\be
 F^{4m}_{i\,j\,k\,\ll} = \Li_2(\alpha_+) +\Li_2(1-\alpha_-)-\Li_2(1)-\log \alpha_+\log(1-\alpha_-) +\frac12\log u\log v \label{fourmassB}
\ee
where $\alpha_\pm = \frac12(1+u-v\pm\sqrt{(1-u-v)^2-4uv})$ and $u=u_{i\,j\,k\,\ll}$ and $v=u_{j\,k\,\ll\,i}$.

Notice, that all expressions are finite and manifestly dual-conformal invariant.
The limit to two dimensions is perfectly smooth in all cases, and the box functions simplify significantly. The square root factorizes and $\alpha^\pm$ become simple cross-ratios,
and $\tilde F^{2me}\to 0$ if $i$ and $j$ have different parity.

Now we study the \Roneone reduction of the box expansion of one-loop 8-point N${}^2$MHV ratio function, which can be expanded using the modified box functions (below we write $F$ instead of $\tilde{F}$),\be
\tilde{R}^{(1),D=4}_{8,2}=-\sum_{i,j,k,\ll} c^{i\,j\,k\,\ll} F_{i\,j\,k\,\ll}\,,\ee
where $c^{i\,j\,k\,\ll}$ are corresponding box-coefficients \footnote{To define the ratio function, one has to subtract the tree amplitude for coefficients of two-mass-easy and one-mass boxes. Explicit expressions of all box coefficients can be found in~\cite{Bourjaily:2013mma}.}, and the summation is over the following boxes: $F^{4m}_{1 3 5 7}$, $ F^{4m}_{2 4 6 8}$, three length-8 cyclic groups $\{F^{3m}_{1 2 4 7},...\}$, $\{F^{3m}_{1 2 4 6},...\}$, $\{F^{3m}_{1 2 5 7},...\}$, three cyclic groups of length-4, $\{F^{2me}_{1 2\,i{-}1\,i},...\}$ for $i=5,6,7$, two-mass-hard and one-mass boxes $\{F_{1 2 3 i},...\}$ for $i=4,...,8$.

When reduced to two dimensions, all box functions only depend on two cross-ratios, $v_1$ and $v_2$. 
As promised, the square root in four-mass boxes disappear: in both $F^{4m}_{1\,3\,5\,7}$ and $F^{4m}_{2\,4\,6\,8}$, $\{\alpha^{\pm}\}=\{\frac {v_1}{1{+}v_1}, \frac{v_2}{1{+}v_2}\}$ (depending on whether $v_1> v_2$ or $v_1<v_2$). The result when $v_1> v_2$ is\ba
&&F^{4m}_{1 3 5 7}=L(v_2)-L(v_1)+\frac12\log(v_1)\log(1{+}v_2)-\frac12\log(v_2)\log(1{+}v_1)\,,\nl
&&F^{4m}_{2 4 6 8}=-L(v_1)-L(v_2)+\frac12\log(v_2)\log(1{+}\frac1{v_1})+\frac12\log(v_1)\log(1{+}v_2)\,,
\ea
and $v_1\leftrightarrow v_2$ for $v_1<v_2$. Here and in the rest of the paper,
\be
 L(x) := \Li_2(-x)+\frac 12\log(x)\log(1{+}x)+\frac{\pi^2}{12},
\ee
which satisfy $L(\frac1 x)=-L(x)$, so it behaves nicely under a cyclic shift by two. All other boxes also simplify a lot, and the two-dimensional octagon can be written as a combination of simple functions such as $L(v_1)$ and $\log (v_1)\log(1+v_2)$, with coefficients reduced from the four-dimensional box coefficients. Since these coefficients are R-invariants, it is important to also reduce the Grassmann variables when taking the \Roneone limit. Although the (supersymmetrically taken) limit for a single box coefficient depends on how it is taken, the coefficient of any independent function in two dimensions must have a smooth and well-defined \Roneone limit since the amplitude itself must be smooth.

For N${}^2$MHV octagon in two dimensions, there is only one independent R-invariant, $R_{8,2}^{\textrm{tree}}=(1\,3\,5\,7)[2\,4\,6\,8]$. Thus, in the limit, any coefficients of the functions above must be proportional to the tree amplitude, with a factor which can be a rational function of cross-ratios. Our task in the reduction of box coefficients is to determine such rational functions.

The most convenient way to obtain the desired limit is to pick a component, such that all box coefficients can have an unambiguous limit for that component. Since there is an unique fermionic combination, $R_{8,2}^{\textrm{tree}}$, one appropriate component is enough to determine the complete result. It turns out both four-mass box coefficients have well-defined \Roneone limits for the component $\chi^1_1\chi^1_3\chi^2_5\chi^2_7\chi^3_2\chi^3_4\chi^4_6\chi^4_8$.  Define $R|_{\text{comp}}:=\l 3\,4\,5\,6\r\l 7\,8\,1\,2\r R|_{\chi^1_1\chi^1_3\chi^2_5\chi^2_7\chi^3_2\chi^3_4\chi^4_6\chi^4_8}$, where the normalization factor that goes to $\l3 5 \r\l 7 1\r[2 4][6 8]$ in the two-dimensional limit. We find that for $v_1>v_2$ (the discussion for $v_1<v_2$ is parallel),
\be c^{1 3 5 7}|_{\textrm{comp}}\to \frac{v_1+v_2}{v_1-v_2},\quad c^{2 4 6 8}|_{\textrm{comp}}\to \frac{2}{v_1 v_2-1},\ee
 Furthermore, it is very convenient that for this component, all other box coefficients reduce to $\pm1$ or 0! The upshot is that, for this component, we find a nice limit of the box expansion, \be \tilde{R}^{(1)}_{8,2}|_{\textrm{comp}}\to -\frac{v_1+v_2}{v_1-v_2} F^{4m}_{1 3 5 7}-\frac{2}{v_1 v_2-1}F^{4m}_{2 4 6 8}+F^{2me}_{3 4 7 8}+F^{2me}_{4 5 8 1}+\ldots\,,\ee
where \ldots represents 8 two-mass easy boxes that vanish in the limit, 8 Li${}_2(1)$'s from two-mass hard boxes (with a minus sign), and 4 from one-mass boxes, so it evaluates to $-\frac{2\pi^2}{3}$.

Plugging the four-mass and two-mass-easy boxes into the combination, we immediately recognize that the result can be organized as a sum of four cyclically-related objects, coming with rational prefactors $\frac{v_1}{v_1-v_2}$, $\frac{v_1 v_2}{v_1 v_2-1}$, $\frac{v_2}{v_2-v_1}$ and $\frac 1 {1-v_1 v_2}$, giving the end result \footnote{Here we have added also the constant corresponding to $c'_4$ in ref.~(\ref{BDS}) to get our BDS-subtracted amplitude from the ratio function, $R^{(1)}_{8,2}=\tilde{R}^{(1)}_{8,2}+ \frac{2 \pi^2}3$. Thanks to this constant, eq.~(\ref{8ptN2MHV})
has the correct $k$-preserving and $k$-decreasing collinear limits $R^{(1)}_{8,2}\to 0$, as can be verified.}:
\be
 R^{(1)}_{8,2} =
 R_{8,2}^\textrm{tree}\frac{v_1}{v_1-v_2} \left(L(v_1)-L(v_2)+\log (1+\frac1{v_1})\log(1+v_2) +\frac{\pi^2}{12}\right) + \mbox{(3 cyclic)}.
\label{8ptN2MHV}\ee
Although this analytic formula was not easy to come by, its correctness can be readily verified numerically using the Mathematica
package provided in ref.~\cite{Bourjaily:2013mma}.

\subsection{Uplifting and the N${}^2$MHV decagon}\label{sec:10ptN2MHV}

We expect that by ``uplifting'' the octagon a significant part of the higher-point amplitudes should be accounted for, although we do not necessarily expect to get the complete answer.  To uplift $R_{8,2}^{(1)}$ we have to ``split'' into two parts $S_{8,2}$ as in (\ref{matching8point}).  Cursory examination of the decagon obtained from four dimensions reveals that the terms multiplying $(1\,3\,5\,7)[2\,4\,6\,8]$ can have $1/(v_1-v_2)$ denominator but not $1/(1-v_1 v_2)$.
Thus we should require $S_{8,2}$ to only have denominators of the form $1/(v_1-v_2)$. Imposing this almost uniquely fixes $S_{8,2}^{(1)}(1,2,3,4,5,6,7,8)$ we have
\be
S_{8,2}^{(1)}= (1\,3\,5\,7)[2\,4\,6\,8]\left[
 \frac{v_1}{v_1-v_2} \left(L(v_1)-L(v_2) +\log (1+\frac1 {v_1})\log(1+v_2) +\frac{\pi^2}{12}\right) + (v_1\leftrightarrow v_2)
 \right]\,.  \label{S82}
\ee
It is trivial to see that $R_{8,2}^{(1)}=S_{8,2}^{(1)}(x_2,x_4,x_6,x_8)+S_{8,2}^{(1)}(x_1,x_3,x_5,x_7)$. Amazingly, as we will show immediately, this also automatically accounts for all the terms in the 10-point amplitude which have a nontrivial denominator!

This makes it algorithmically very easy to extract the remainder of the 10-point amplitude, by taking semi-numerically the two-dimensional limit of the coefficients of the various transcendental functions.  Algorithmically speaking, this would be a difficult task if these coefficients were to be complicated rational functions of the external momenta.  But thanks to the fact that these coefficients are now simple integers (or half integers), the problem becomes easy.  One just has to evaluate the coefficients numerically.

In this way we could obtain the full N${}^2$MHV decagon in analytic form:  \ba R^{(1)}_{10,2}=\sum_{1\leq i\lhd j\lhd k\lhd \ll\leq 10}
(-)^{i{+}j{+}k{+}\ll} S_{8,2}^{(1)}(x_i,x_j,x_k,x_{\ll})+r_{10,2}^{(1)}. \label{10ptN2MHV} \ea Here we have a sum of  25 different
$S_{8,2}^{(1)}$ terms, which can be classified into three cyclic classes explicitly:
\begin{itemize}
\item $(i,j,k,\ll)$=(2,4,6,8) and 9 cyclic:  $(1\,3\,5\,7)[2\,4\,6\,8]f_{8,2}(v_1,v_2)+ \mbox{(9 cyclic)}$;
\item $(i,j,k,\ll)$=(1,4,6,8) and 9 cyclic:  $-(1\,3\,5\,7)[4\,6\,8\,10]f_{8,2}(\frac 1 {v_1}, v_4)+ \mbox{(9 cyclic)}$;
\item $(i,j,k,\ll)$=(1,3,6,8) and 4 cyclic: $(1\,3\,5\,7)[2\,6\,8\,10]f_{8,2}(\frac 1 {v_1}, \frac 1 {v_6})+\mbox{(4 cyclic)}$.
\end{itemize}
where $f_{8,2}(v_1,v_2)$ stands for the bracket in (\ref{S82}). The remainder, $r_{10,2}^{(1)}$, only contains transcendental degree-two
functions with coefficients given by pure R-invariants (\emph{i.e.} no rational prefactors; its expression is given in the
Appendix~\ref{app:r10}. It is remarkable that the uplifting accounts for all non-trivial contributions to the 10-pt amplitude, \emph{i.e.} terms
with rational prefactors.

Finally we note that the expression (\ref{S82}) is not uniquely determined, but since $S_{8,2}^{(1)}$ must be invariant under cyclic rotations by two,
there is only one ambiguity which is a multiple of $\log v_1\log v_2$.  This is odd under rotations by 1, and so will not contribute to $R_{8,1}$. However, we found that adding such a term does not lead to any simplification of the ``remainder'' $r_{10}$, so we didn't add it.

Although we didn't include the explicit formula here, it is straightforward to compute also the 12-point N${}^2$MHV amplitude, which we will use
later to generate 10-point NMHV at two-loop from the $\bar Q$ equation. As we checked arithmetically, the uplifting of $S_{8,2}^{(1)}$ again removes all non-trivial terms (\emph{i.e.} those with rational prefactor) for the 12-point amplitude. Moreover, as conjectured in \cite{Goddard:2012cx}, we expect the information of 10-point and 12-point remainders should be enough to uplift $n$-point one-loop N${}^2$MHV amplitudes.

\section{The $\bar Q$ equation and the two-loop NMHV octagon}\label{2loopNMHV}

We now return to the main object of this paper, which is to study the higher-loop MHV and NMHV octagons.
Out of the equations of \cite{CaronHuot:2011kk} encoding integrability,
we will use only the $\bar Q$ equation, which admits a reduction to the two-dimensional subsector
considered in the present paper. (The other equation, related to the $S^{(1)}$ symmetry, does not close within that subsector.)
The $\bar Q$ equation by itself does not encode integrability, only Poincar\'e supersymmetry of the dual Wilson loop,
but as shown in \cite{CaronHuot:2011kk} it is still extremely powerful when applied to MHV and NMHV amplitudes.

When reduced to two-dimensional kinematics, the $\bar Q$ equation takes the form \cite{CaronHuot:2011kk}
\ba
\hspace{-0.3cm}
 \bar Q^A_a R_{n,k}  = \aa \int d^{1|2}\lambda_{n{+}1}\int d^{0|1}\lambda_{n{+}2} (R_{n{+}2,k{+}1} -
  R^\textrm{tree}_{n{+}2,1} R_{n,k}) + \textrm{cyclic}  \label{Qbar}
\ea
where the cusp anomalous dimension, $\aa=g^2-\frac{\pi^2}3 g^4+\ldots$ (with $g^2\equiv\frac{g^2_\textrm{YM}N_c}{16\pi^2}$), is chosen to be our loop expansion parameter.
The right-hand side involves three fermionic integrations, consistent with the quantum numbers of $\bar Q$, as well as one bosonic integration.
The new twistors $\lambda_{n{+}1}$ and $\lambda_{n{+}2}$ are added in such a way as to effectively insert an infinitesimal ``kink'' along edge $n$,
as detailed shortly.  The cyclic sum covers the even edges of the polygon, onto which the kink is to be inserted.

We discuss, in turns, the left-hand-side and right-hand-side of this equation, and apply it to obtain the \emph{two-loop}
NMHV octagon starting from the one-loop N${}^2$MHV decagon obtained in the previous section.

\subsection{Left-hand side of the $\bar Q$ equation}\label{LHS}

The left-hand side of the $\bar Q$ equation is a first order differential operator
\be
 \bar Q^{A}_\alpha R_{n,k} = \sum_{\substack{i=2\\ \textrm{$i$ even}}}^{n}
 \chi_i^A \frac{\partial}{\partial Z_i^\alpha} R_{n,k}\,,
\ee
where both the R-symmetry and twistor indices $A$ and $\alpha$ belong to the even sector.
(There is a similar equation for the odd sector, which we will not require.)

For concreteness let us consider the case $n=8$, $k=1$.  The left-hand-side involves
$R_{8,1}$ which is given by R-invariants times bosonic functions of cross-ratios, which are parametrized in (\ref{R81}).
Since $\bar Q$ acts only on even sector variables, let us ignore for a moment
the dependence on the odd sector variables.
Due to the unbroken $Q$ and bosonic dual conformal symmetries, the parametrization takes the form
\be
 R_{8,1} = [2\,4\,8]F_1(v_2)  + [2\,6\,8]F_2(v_2) + [4\,6\,8]F_3(v_2), \label{paramR81}
\ee
where $v_2$ is the independent cross-ratio in the even sector.
Since $\bar Q$ commutes with the R-invariants (by construction) it is effectively a derivative with respect to $v_2$.
To extract useful information from it we need the fact that $\bar Q$ has two twistor indices, \emph{and} two distinct (even) twistor components:
\be
 \lambda_4^\alpha \bar Q^A_\alpha v_2 = v_2\frac{[4\,8]\l\l 6\,8\,2\r\r^A}{[6\,8][8\,2]},
 \qquad
   \lambda_6^\alpha \bar Q^A_\alpha v_2 = v_2\frac{[2\,6]\l\l 8\,2\,4\r\r^A}{[8\,2][2\,4]}
 \label{dweq1}
\ee
where $\l\l i\,j\,k\r\r:=[i\,j]\chi_k+[j\,k]\chi_i+[k\,i]\chi_j$.
Consider just the first of these components. Using that $[2\,6\,8]\l\l 2\,6\,8\r\r=0$ its action on (\ref{paramR81}) gives
\be
\lambda_4^\alpha \bar Q^A_\alpha R_{8,1} = v_2([2\,4\,8]F_1'(v_2)+[4\,6\,8]F_3'(v_2))\frac{[4\,8]\l\l 6\,8\,2\r\r^A}{[6\,8][8\,2]}.\label{dweq2}
\ee
The crucial point is that both $F_1'$ and $F_3'$ appear in this equation.  Furthermore
these can be extracted independently from different Grassmann components.  Specifically,
setting $A=3$, the component $\chi_2^3\chi_4^3\chi_2^4$ selects out $F_1'(v_2)$ while $\chi_2^3\chi_4^3\chi_6^4$
selects out $F_3'(v_2)$.  All other components are related to these two by supersymmetry.

Since the equation for $\lambda_6^\alpha Q_\alpha^A$ similarly allows to extract $F_2'(v_2)$ and $F_3'(v_2)$ separately,
we conclude that the left-hand-side of the $\bar Q$ equation gives the derivative of all bosonic functions entering $R_{8,1}$.
This explains the usefulness of the $\bar Q$ equation at NMHV level.

The attentive reader will have noticed that the above gives two distinct ways to extract $F_3'(v_2)$, from $\lambda_4^\alpha\bar Q_\alpha$ or $\lambda_6^\alpha\bar Q_\alpha$.
In other words, the system is over-constrained. That is, the left-hand-side of the $\bar Q$ equation automatically satisfies
\be
[2\,6](\lambda_4^\alpha \bar Q^A_\alpha R_{8,1})\big|_{\chi_2^3\chi_4^3\chi_6^4} - [2\,4](\lambda_6^\alpha \bar Q^A_\alpha R_{8,1})\big|_{\chi_2^3\chi_6^3\chi_4^4} =0\,. \label{nontrivial0}
\ee
Ultimately this must be viewed as a constraint on the right-hand side of the $\bar Q$ equation, e.g. as a nontrivial constraint on the
lower-loop amplitude which enters it. We will see that it is satisfied by the one-loop amplitudes obtained previously.

We can now restore the dependence on the odd variables, and by solving a simple linear algebra problem we find the relation between
the components of the $\bar Q$ equation and the derivatives of the functions $f_{8,1}^{i}$ with $i=1,2$ entering eq.~(\ref{R81}):
\ba \hspace{-0.7cm}
 \frac{\partial}{\partial v_2} f_{8,1}^{1}(v_1,v_2) &=& \frac{\l1\,3\r[2\,6]}{1+v_2}\left( \lambda_8^\alpha\bar Q^3_\alpha R_{8,1}
  \big|_{(\chi_1^1\chi_3^2)(\chi_2^3\chi_8^3\chi_6^4)}\right)
\nl\hspace{-0.7cm}
\frac{\partial}{\partial v_2} f_{8,1}^{2}(v_1,v_2) &=& \frac{\l1\,5\r[2\,4]}{v_2}\left(
\lambda_6^\alpha\bar Q^3_\alpha R_{8,1}\big|_{(\chi_1^1\chi_5^2)(\chi_4^3\chi_6^3\chi_2^4)}
+2\frac{\l1\,3\r}{\l1\,5\r} \frac{\lambda_6^\alpha\bar Q^3_\alpha R_{8,1}}{1+v_2}\Big|_{(\chi_1^1\chi_3^2)(\chi_2^3\chi_6^3\chi_4^4)}
\right)\!.
 \label{combinddu2}
\ea

\subsection{Right-hand side of the $\bar Q$ equation}\label{RHS}

We must now evaluate the right-hand side of the $\bar Q$ equation (\ref{Qbar}).
For $n=8, k=1$ this requires a 10-point N${}^2$MHV amplitude, but at a lower loop order.
Thus the one-loop results of the previous section will allow us to obtain the two-loop NMHV octagon.

Before we do this, it is instructive to review the computation of the one-loop NMHV octagon starting from the tree amplitude on the right-hand
side.  This computation was performed in \cite{CaronHuot:2011kk} and our aim here is to reproduce it with greater detail.

In the even sector, the $d^{0|1}$ operation appearing in (\ref{Qbar}) is defined by setting $z_{n+2}=z_n + \epsilon \frac{[n\,4]}{[2\,4]} z_2$
supersymmetrically (\emph{after} performing the Grassmann integration), followed by taking the soft-collinear limit $\epsilon\to 0^+$.
On simple R-invariants this gives
\be
 \int d^{0|1}\lambda_{n{+}2} [i\,n{+}2\,j] = \lambda_n \frac{\l\l i\,n\,j\r\r}{[i\,n][n\,j]} =\bar Q \log \frac{[i\,n]}{[j\,n]}, \quad i\neq n, \label{d0slash1}
\ee
with $[i\,n]$ replaced by $[2\,n]$ when $i=n$. On R-invariants not containing $z_{n{+}2}$, the operation gives zero because of the Grassmann integration.
(Note that these considerations also imply that $[2\,n\,n{+}2]$ is mapped to zero.)

In the odd sector there is no limit involved, and one simply integrates over
the two Grassmann components of $z_{n{+}1}$ as well as over its bosonic components (which only has one degree of freedom due to projective
invariance). Thus on simple R-invariants one gets
\be
 \int d^{1|2} \lambda_n{+}1\, (i\,j\,n{+}1) = \int_{\lambda_{n{+}1}=\lambda_{n{-}1}}^{\lambda_{n{+}1}=\lambda_1} d\log \frac{\l i\,n{+}1\r}{\l j\,n{+}1\r}, \label{d1slash2}
\ee
the result being zero for $R$-invariants not containing ($n{+}1$).
The integration runs over the position of the ``kink'' $\lambda_{n{+}1}$, which is inserted along segment $n$.

These formulas are all we need to compute the right-hand-side of the $\bar Q$ equation.
Using eq.~(\ref{n2mhvtree}) for the N${}^2$MHV tree amplitude, keeping only the non-vanishing terms,
\be
 R^\textrm{tree}_{10,2} \simeq (3\,5\,7)(3\,7\,9)[4\,6\,8][4\,8\,10]+(1\,3\,5)(5\,7\,9)[2\,4\,10][6\,8\,10]\,,
\ee
and plugging (\ref{d0slash1}) and (\ref{d1slash2}) into it, 
one obtains the contribution from edge 8:
\be\hspace{-0.1cm}
 \bar Q R_{8,1} \big|_{\textrm{edge 8}} \supset
 \bar Q \log \frac{[6\,8]}{[2\,8]}\int_{\lambda_9=\lambda_7}^{\lambda_9=\lambda_1} \left(
 (3\,5\,7)[4\,6\,8] d\log \frac{\l7\,9\r}{\l 3\,9\r}
 + (1\,3\,5)[2\,4\,8]d\log \frac{\l7\,9\r}{\l 5\,9\r}
 \right). \label{r81edge8}
\ee
Notice that this is logarithmically divergent near the endpoint $\lambda_9= \lambda_7$.  However, eq.~(\ref{Qbar}) requires us to subtract a second term \emph{under} the integration sign, related to the $\bar Q$ variation of the BDS Ansatz, which is itself divergent.  The subtraction gives
\be
 -R_{8,1}^\textrm{tree} \int_{\lambda_9=\lambda_7}^{\lambda_9=\lambda_1}\left(
  \bar Q \log\frac{[4\,8]}{[2\,8]} d\log \frac{\l 3\,9\r}{\l 5\,9\r}
  - \bar Q\log \frac{[6\,8]}{[2\,8]}d\log \frac{\l 7\,9\r}{\l5\,9\r} \right)  \label{r81subs}
\ee
with
\be
 R_{8,1}^\textrm{tree} = (3\,5\,7)[2\,6\,8]-(1\,5\,7)[2\,6\,8]-(1\,3\,5)[2\,4\,8]-(3\,5\,7)[4\,6\,8]\,.
\ee
Miraculously, using the identity $[2\,6\,8]\bar Q\log \frac{[6\,8]}{[2\,8]}=0$, one finds that the divergent terms involving $d\log \l 7\,9\r$ cancel precisely between the two expressions.

With hindsight, this ``miracle'' is not too surprising since the relevant integration region corresponds to a soft limit of the 10-point scattering
amplitude, where the cancelation is expected to occur at any loop order due to factorization~\cite{CaronHuot:2011kk}.  The fact that the
divergences cancels separately for each edge is as a useful consistency check.

Thus the sum of (\ref{r81edge8}) and (\ref{r81subs}) is a finite integral, which we can simplify to
\be
\bar Q R_{8,1} \big|_{\textrm{edge 8}} = \big[ (3\,5\,7)-(1\,5\,7)\big][2\,6\,8]\bar Q\log \frac{[4\,8]}{[2\,8]}
\int_{\lambda_9=\lambda_7}^{\lambda_9=\lambda_1}d\log \frac{\l 5\,9\r}{\l 3\,9\r}\,. \label{intw0}
\ee
The integration is elementary and gives $\log(1+\frac{1}{v_1})$.  We note that by using a covariant parametrization,
such as $\lambda_9=\lambda_7 + \tau \frac{\l 7\,3\r}{\l 1\,3\r} \lambda_1$ with $\tau>0$,
the integrand itself would be directly a function of cross-ratios.

To obtain the right-hand side of the $\bar Q$ equation (\ref{Qbar})
it remains to sum over all even edges of the octagon. This gives four terms like (\ref{intw0}), all
related by symmetry.  It is possible to verify that the constraint (\ref{nontrivial0}) is satisfied,
thanks to nontrivial cancelations among all four edges.
Therefore we are ready to evaluate (\ref{combinddu2}).  Amusingly, we find that $\partial/\partial v_2 f_{8,1}^{1(1)}$ comes entirely from edge 6.
Explicitly it is given as
\ba
 \frac{\partial}{\partial v_2} f_{8,1}^{1(1)}(v_1,v_2) &=& \frac{\l1\,3\r[2\,6]}{1+v_2}
 \left.\big[(1\,3\,5)-(3\,5\,7)\big]\log(1+v_1)[4\,6\,8](\lambda_8^\alpha\bar Q^3_\alpha\log \frac{[2\,6]}{[6\,8]})
 \right|_{(\chi_1^1\chi_3^2)(\chi_2^3\chi_8^3\chi_6^4)}
 \nl
 &=&\frac{1}{1+v_2} \log(1+v_1)\, . \label{intw1}
\ea
The solution is $f_{8,1}^{1(1)}(v_1,v_2)=\log(1+v_1)\log(1+v_2)$,
up to an integration constant. The latter is set to zero by the collinear-soft limit conditions.
One can proceed similarly for $f_{8,1}^{2(1)}$, and the final result is
\be
 f_{8,1}^{1(1)} = \log(1+v_1)\log(1+v_2), \qquad
 f_{8,1}^{2(1)} = \log \frac{(1+v_1)^2}{v_1} \log \frac{(1+v_2)^2}{v_2} \,
\ee
exactly as claimed in section \ref{octagonNMHV}, where the results of~\cite{CaronHuot:2011kk} (which used the same $\bar Q$ equation as here) were reported.

\subsection{Two-loop NMHV octagon}\label{8ptNMHV}

By inserting the one-loop decagon (\ref{10ptN2MHV}) into the right-hand side of the $\bar Q$ equation, the same procedure will give
the derivative of the two-loop NMHV octagon.
The algebraic steps of the previous subsection, and in particular, eqs.~(\ref{combinddu2}), are unchanged.
However the $\lambda_9$ integrand in (\ref{intw0})
is now a weight 2 transcendental function inherited from the decagon, so the integration step is less easy.

It is still possible to (non-canonically) lump all edge contributions under a common integration sign, and extract
the components (\ref{combinddu2}). These steps involve simple linear algebra
and the transcendental functions carry along for the ride.
Thus one obtains an integro-differential equation of the form
\be
  \frac{\partial}{\partial v_2} f_{8,1}^{i(2)}(v_1,v_2) = \int_0^\infty d\tau F^i(v_1,v_2,\tau) \label{diff2loop}
\ee
where the $F^i$ are explicitly known transcendental functions of weight 2 inherited from the one-loop decagon, times rational factors.
The expressions are too lengthy to be reported here, although we found that the integrations could be performed numerically very easily.

The integrations cannot be done analytically by elementary methods (\emph{i.e., Mathematica}). But they can be done immediately at the level of
the so-called symbol, using simple linear algebra following the algorithm of Appendix A of ref.~\cite{arXiv:1105.5606}. In this way (and using
symmetry to obtain derivatives with respect to $v_1$) one obtains the symbol of the desired weight 4 functions. One then faces the problem of
finding corresponding functions. Since the relevant procedure is technical and unspecific to the present problem, we postpone its discussion to
the next section.

The upshot is that with the help of symbol technology, eq.~(\ref{diff2loop}) can be integrated analytically.
Here we give the result, which is expressed in terms of the multiple polylogs
\be
 \Li_{i_1,i_2,\ldots,i_k}(x_1,x_2,\ldots,x_k) := \sum_{a_1>a_2>\ldots>a_k\geq 1} \frac{x_1^{a_1}}{a_1^{i_1}}\frac{x_2^{a_2}}{a_2^{i_2}} \cdots \frac{x_k^{a_k}}{a_k^{i_k}}.
\ee
Writing $(v_1,v_2)\mapsto (v,w)$ for better readability, the result is
\ba
f^{1(2)}_{8,1} &=& -2\Li_{2,2}(-\frac 1 w,vw)-2\Li_{1,3}(-\frac 1 w,vw)-4\Li_4(-v)+\log \frac{v}{w}\left[\Li_{1,2}(-\frac 1 w,vw)+\Li_3(-v)\right]
\nl &&
+ 6\log(1+v)\Li_3(-w)-\big[\Li_2(-v)+2\log(1+v)\log w\big]\big[\Li_2(-w)+\log vw\log(1+w)\big]
\nl &&
+\log(1+w)\left[4\Li_3(-v)+ \log(1+v)\left(\log \frac{(1+v)^2}{v}\log(1+w)-\log \frac{(1+v)}{(vw)^2}\log w+\frac{2\pi^2}{3}\right)\right],
\nl
f^{2(2)}_{8,1} &=& 
\left[ 6\Li_3(-w)-2\Li_2(-w)\log w+\frac{\pi^2}{6}\log w(1+w)^2+\frac18\log^2w\log \frac{(1+v)^2}{v}\right]\log \frac{(1+v)^2}{v}
+ (v\leftrightarrow w)
\nl &&
  + \frac38\left[\log^2\frac{(1+v)^2}{v}\log^2\frac{(1+w)^2}{w} -\log^2v\log^2w\right]
 + \frac{\pi^2}{12}(\log^2v+\log^2w) + \frac{\pi^4}{20}.  \label{mainresult2loop}
\ea
The first of these functions is symmetrical under $v\leftrightarrow w$, although this is not manifest from the expression.
Similarly, the second function is invariant under the full dihedral group. (The explicit symmetrization $(v\leftrightarrow w)$ in the first line applies to that line only.)


A few comments are in order.
\begin{itemize}
\item We verified that the constraint (\ref{nontrivial0}) was satisfied; this can be viewed as a nontrivial property of the one-loop decagon.
\item The $\bar Q$ equation can be used to produce separately the derivatives $\partial_{v_1} f$ and $\partial_{v_2}f$.
It is not \emph{a priori} manifest that these will be consistent, \emph{e.g.} that $\partial_{v_2} (\partial_{v_1} f) = \partial_{v_1}(\partial_{v_2} f)$, but we find that this is the case thanks to a remarkably nontrivial interplay between the various edge contributions.
We view this as another nontrivial property of the decagon.
\item Although we have used the symbol map in intermediate steps, and the symbol map is lossy, in the end our result is unambiguous
because we could verify numerically the differential equation (\ref{diff2loop}).  The (single) otherwise undetermined integration constant which was determined using numerics is described in section \ref{sec:method}.
\item The $\bar Q$ equations are first-order differential equations and so determine $f^{1}$ and $f^2$ only up to constants.
These constants are fixed by the collinear-soft limits. (These limits are particularly simple thanks to our non-standard renormalization constants in eq.~(\ref{BDS}), which trivialize the 6-point amplitudes in \Roneone kinematics.)
\end{itemize}

Let us verify explicitly the soft-collinear limits,
\ba
 0&=&\lim_{v\to0} f^{1(2)}_{8,1}(v,w),\nl
 0&=&\lim_{w\to\infty} \left(f^{1(2)}_{8,1}(v,w)+ f^{1(2)}_{8,1}(1/v,w) -f^{2(2)}_{8,1}(v,w)\right).
\ea
The first limit is easily verified in eq.~(\ref{mainresult2loop}) --- every term in the expression vanishes.
The second limit requires a little more work.  The $\Li_{i,j}$ terms vanish in the limit,
and a simple computation gives
\ba
 \lim_{w\to\infty} f_{8,1}^{1(2)}(v,w) &=& \phantom{+}\log^2(w)\big[\log(1+v)\log(1+1/v) -\frac12\Li_2(-v)\big]
 \nl &&+ \log^1(w)\big[3\Li_3(-v)-\log v\Li_2(-v)\big]
 \nl &&+ \log^0(w)\left[-4\Li_4(-v)+\Li_3(-v)\log v+\frac{\pi^2}{6}\Li_2(-v)\right] + \mathcal{O}(1/w)\,.\nl
\ea
This simplifies a bit upon adding the term with $v\mapsto 1/v$; in particular the $\Li_4$ term disappears thanks to a polylog identity.
The result cancels precisely the limit of $f_{8,1}^{2(2)}$.

\subsection{Uplifting for two-loop NMHV}\label{10ptNMHV}
\label{sec:upliftNMHV}

Emboldened by the success of the uplifting formalism at one-loop for both NMHV and N${}^2$MHV, it is natural to turn our attention to the decagon.
Our strategy was to generate its symbol at two-loop using the $\bar Q$ equation and 12-point N${}^2$MHV, which is entirely similar to the
preceding computation, and then try to see if it could be ``mostly'' understood as some uplifting of 8-point.

The symbol of $R_{8,1}^{(2)}$ contains ``nontrivial'' entries of the form $v_1-v_2$ and $1-v_1v_2$; physically these represent branch points of the amplitude on some higher Riemann sheet. (The third sheet, to be precise, as these appear only in the third entry of the symbol.)  These combinations appeared previously as poles of the N${}^2$MHV \emph{one-loop} amplitude.
In that context we found in \mbox{section \ref{sec:10ptN2MHV}} that the correct uplifting $S_{8,2}^{(1)}$ included terms with only the first type of pole, $v_1-v_2$.
It is natural to try the same here, \emph{e.g.} uplift all terms which have $v_1-v_2$ in their symbol, e.g.
the contributions $f_{8,1}^{1(2)}(1/v_1,v_2)$ and $f_{8,1}^{1(2)}(v_1,1/v_2)$ in eq.~(\ref{mainresult2loop}).
Remarkably, this simplest guess turns out to reproduce all terms the decagon symbol containing $(v_\textrm{odd}-v_\textrm{even})$-type entries!

Having thus convinced ourselves that the 2-loop NMHV amplitude ``wants'' to be uplifted, we set out to do it in a systematic fashion.
The most general Ansatz for $S_{8,1}(1,2,3,4,5,6,7,8)$ takes the form
\ba
 S_{8,1} &=& \phantom{+}\big((1)[2]+(5)[6]\big) g_1(v_1,v_2)+ \big( (3)[4] + (7)[8] \big) g_1(\frac 1{v_1},\frac{1}{v_2})
 \nl && +\big( [2](3)+ [6](7) \big) g_2(v_1,v_2) +\big( [4](5)+[8](1)\big) g_2(\frac 1 {v_1},\frac 1 {v_2})
\nl  && + \big( (1)+(5)\big)\big([2]+[6]\big) g_3(v_1,v_2). \label{eightpointNMHVS}
\ea
This is similar to (\ref{eightpointNMHV}) for $R_{8,1}$, except that we have changed the last line to something that is more appropriate to the symmetries of $S_{8,1}$.
To convince oneself that covers is the most general $S_{8,1}$, one notes that the 8 products of R-invariants in the first two lines, together with the last line, span a basis of 9 linearly independent objects at 8-points; thus any Ansatz can be cast into this form.

Now, $S_{8,1}$ must be invariant under interchange of odd and even labels. This forces $g_1(v,w)=g_1(w,v)$ but \emph{also}
$g_2(v,w)=-g_2(w,v)$ --- the R-invariants on the second line transform among themselves in a nontrivial way.
These symmetry properties ensure that the functions $g_1$ and $g_2$ are uniquely determined by $R_{8,1}$, while there is a slight ambiguity in
$g_3$ since only the sum $g_3(v,w)+g_3(1/v,w)$ enters $R_{8,1}$.
Furthermore, since the formalism is built to trivialize collinear limits, it is more natural to apply it to the two-loop
\emph{ratio} function $R_{8,1}^{(2)}-R_{8,1}^{\textrm{tree}}R_{8,0}^{(2)}$ where $S_{8,0}^{(2)}=-\log(1+v)\log(1+1/v)\log(1+w)\log(1+1/w)$.
Thus one finds
\ba
 S^{(2)}_{8,1} &=& \phantom{+} \big( (1)[2]+(3)[4]+(5)[6]+(7)[8] \big) \frac12\left(f_{8,1}^{1(2)}(v,1/w) + f_{8,1}^{1(2)}(1/v,w) -f_{8,1}^{2(2)}(v,w)+S_{8,0}^{(2)}\right)
  \nl && + \big( [2](3)-[4](5)+[6](7)-[8](1)\big) \frac12\left(f_{8,1}^{1(2)}(v,w) - f_{8,1}^{1(2)}(1/v,1/w)\right)
\nl && + \big( (1)+(5)\big)\big([2]+[6]\big) \left( \frac14f_{8,1}^{1(2)}(v,w) -\frac12S_{8,0}^{(2)}+ \delta_{8,1}(v,w)\right)\,. \ea
There is a possible ambiguity which sits in the last term and has the following symmetries:
$\delta_{8,1}(v,w)=\delta_{8,1}(w,v)=-\delta_{8,1}(1/v,w)$.  Since we noted that the $f_{8,1}^1$ terms already reproduce all
$(v_\textrm{odd}-v_\textrm{even})$ entries in the higher-point symbols, we do not want to add such entries to $\delta_{8,1}$. Thus we will only
consider ambiguities which are simply products of functions of $v$ and $w$, the class of functions considered in ref.~\cite{Duhr:2012fh}. The
most general such ambiguity at two loops, with the desired symmetries, is parameterized by 7 coefficients: \ba
 \delta_{8,1}^{(2)} &\supset& L(v)L(w),
 \quad L(v)\log w\log \frac{(1+w)^2}{w},
 \quad L(v)\log \frac{(1+v)^2}{v}\log w,
 \quad  \zeta(2)\log v\log w\,,
 \nl &&
 \log v \log^3 w,
 \quad \log v\log w\log \frac{(1+v)^2}{v}\log \frac{(1+w)^2}{w},
 \quad \log v \log w \log^2\frac{(1+w)^2}{w},
 \ea
to be symmetrized in $(v\leftrightarrow w)$.

Remarkably, we find that a unique combination exists for $\delta_{8,1}$ such that
$S^{(2)}_{8,1}$ \emph{does not diverge} in collinear limits:
\ba
 \delta_{8,1} &=& -L(v)L(w)-\frac14\big(L(v)\log w\log \frac{(1{+}w)^2}{w}+L(w)\log v\log \frac{(1{+}v)^2}{v}\big)
 \nl &&+\frac{1}{16}\log v\log w\log \frac{(1{+}v)^2}{v}\log \frac{(1{+}w)^2}{w} -\frac{\pi^2}{12}\log v\log w + \frac{7}{4}\zeta(4)\,. \label{delta81}
\ea
The $\zeta(4)$ constant does not have the required symmetry properties, but was added for reasons to be explained shortly.

Thus we propose the following uplifting formula for general $n$-point:
\be
 R_{n,1}^{(2)} -R_{n,1}^\textrm{tree}R_{n,0}^{(2)} = -21\zeta(4) R_{n,1}^\textrm{tree} +
\sum_{1\leq i< j<k<\ll\leq n} (-)^{i{+}j{+}k{+}\ll} S_{8,1}^{(2)}(x_i,x_j,x_k,x_\ll)+r_{n,1}^{(2)}\,. \label{generalnmhv2}
\ee

Contrary to eq.~(\ref{generaluplift}), note that the summation here includes ``boundary terms'' which depend only on 6 or 7 twistors.
For example for $n=10$ the summation contains 25 off-shell terms where none of the $x_i$ are null separated,
but also 100 simpler ``boundary terms'' which depend only on 7 twistors,
as well as 65 double-boundary terms which depend only on 6 twistors (of which only 50 are nonzero).
The inclusion of these additional terms ensures that the uplifting Ansatz has correct collinear limits, even though $S_{8,1}^{(2)}$ cannot be chosen to vanish in collinear limits. This works thanks to the fact that $S^{(2)}_{8,1}$ has well-defined limits, \emph{and} vanishes when two points become identical $x_{i{+}1}\to x_i$, which turns out to be true for the choice (\ref{delta81}).

The double-boundary terms depending on 6 twistors are simply R-invariants times $\zeta(4)$, but they introduce a new problem since they (incorrectly) make the Ansatz nonzero for $n=6$. This is solved by including a term in the Ansatz proportional to the tree amplitude (which does not yet seem to admit any simple uplifting-type formula \cite{Goddard:2012cx}), and by tuning its coefficient as well as compensating constants in $S_{8,1}^{(2)}$ in a unique way so as to get the correct $n=6$ and $n=8$ results. This explains the $\zeta_4$ terms in (\ref{delta81}) and (\ref{generalnmhv2}).

In summary, eqs.~(\ref{generalnmhv2}) together with eqs.~(\ref{eightpointNMHVS}) and (\ref{delta81}) give one's unique best guess for the $n$-point amplitude, given only the $n=8$ result.

We have computed the ``remainder'' $r_{n,1}^{(2)}$ for $n=10$ and $n=12$ using the same method ($\bar Q$ equation) as for the octagon.  To our amazement, we find that it contains only $\log^4$ terms, not even any constant term or multiple of $\pi^2$!
It would be nice to find a compact closed-form formula for these $\log$ terms, perhaps expressed as the uplifting of some simple $S_{12}$.
The remainders $r_{10,1}^{(2)}$ and $r_{12,1}^{(2)}$ can be found in an ancillary file (following the link) \href{http://www.nbi.dk/~schuot/nmhvremainders.zip}{nmhvremainders}. 

\section{Procedure for integrating symbols}\label{integratesymbol}

The space of two-variables polylogarithms is large and full of identities. Correspondingly, a given function such as
$f^{1(2)}_{8,1}$ can be expressed in terms of $\Li_{i,j}$'s in many different ways; generically the expressions will be much more lengthy than eq.~(\ref{mainresult2loop}). We will now describe the method we have followed, in order to obtain reasonably compact integrated expressions for a given symbol without too much trial and error.
The method also turned out to be successful at three-loops, and hopefully may be applicable in other situations as well.

Methods for integrating symbols have been discussed elsewhere, and our discussion has much overlap with \cite{Duhr:2012fh}. We will try to
emphasize the aspects which are specific to the present problem.

\subsection{Grading by complexity}

Any reasonable integration strategy uses the fact that polylogs form an algebra graded by the weight.
The grading also allows to distinguish among functions of a given weight:
the weight 4 function $\Li_3(x)\log(x)$ is ``simpler'' than $\Li_4(x)$ since it is a product of lower-weight functions.
This notion can be formulated directly at the level of the symbol.

We will not review the definition of symbols here, but refer instead to refs.~(\cite{Goncharov:2010jf}). For completeness, let us only state
that to any weight $k$ transcendental function (within the class of functions, polylogarithms, that are relevant to our problem) is associated a
symbol. This is an element of the $k$-fold tensor product of a (multiplicatively denoted) vector space $\SS_1$, the objects in $\SS_1$ encoding
the possible branch points of the function (in all Riemann sheets). For instance, in the case of classical poylogarithms $\Li_n(x)$,
$\SS_1=\{x,1-x\}$ and the symbol map gives \emph{e.g.} \be
 \SS \Li_4(x) = -\big[(1-x)\otimes x\otimes x\otimes x\big], \quad \frac{1}{4!}\SS \log^4 x = \big[ x\otimes x\otimes x\otimes x\big]\,.
\ee
Interest lies in the fact that all known polylogarithms identities are generated by trivial linear identities on the symbol.
Furthermore the symbol of an integral can often be obtained via simple algebraic operations on its integrand.

There is a standard operation on a symbol which removes all products of lower weight functions (see the review~\cite{Duhr:2012fh}). Let $\SS_k$
denote the vector space of all symbols\footnote{Not necessarily ``integrable.''} of a given length $k$. Given any symbol in $\SS_k$, we define
its $\BB_2\otimes \SS_{k-2}$ projection by antisymmetrizing in the first two entries \be
 a_1\otimes \cdots\otimes a_k\big|_{\BB_2\otimes \SS_{k-2}} := \left[ (a_1\otimes a_2)-(a_2\otimes a_1)\right] \otimes \cdots\otimes a_k\,.
\ee
(This is a projection operator up to a normalization $\frac12$, which we will ignore in the following.)
For $i>2$ we define $\BB_i\otimes \SS_{k-i}$ projectors inductively, as the ``nested commutators''
\be
 a_1\otimes \cdots \otimes a_k \big|_{\BB_i\otimes \SS_{k-i}} :=
 \left.\left[ \big(a_1\otimes a_2\otimes\cdots\otimes a_i\big) - \big(a_2\otimes \cdots\otimes a_i\otimes a_1\big)\right]\right|_{\BB_{i{-}1}\otimes \SS_1}
 \otimes a_{i{+}1}\otimes\cdots \otimes a_k\,.
\ee
Thus, for instance,
\be
 a\otimes b\otimes c\big|_{\BB_3} = (a\otimes b\otimes c)-(b\otimes a\otimes c)-(b\otimes c\otimes a)+(c\otimes b\otimes a)\,.
\ee
The $\BB_i$ projection removes all symbols arising from shuffle products of symbols of weight lower than $i$, formalizing the notion
at the beginning of this subsection.

Starting from weight 4 the classification can be refined further.
The
function
\be
\SS \Li_{2,2}(x,1) = (1-x)\otimes x\otimes (1-x)\otimes x
\ee
is ``more complicated'' than any $\Li_4(\cdots)$ since its $\BB_4$ projection is not symmetrical in the last two entries,
\be
\Li_{2,2}(x,1)\big|_{\BB_4;\BB_2\otimes \BB_2}  = \left[ x\wedge (1-x)\right] \otimes \left[x\wedge (1-x)\right] \neq 0\,.
\ee
The antisymmetrization (related to the so-called ``depth'' of a polylogarithm) kills all $\Li_4$'s, allowing one to concentrate on the depth-two sector.

Within this sector one can further distinguish between functions of one argument (expected to be given by HPLs with a single argument), and genuine functions of two distinct arguments, such as $\Li_{2,2}(-1/w,v w)$.
This can be done by antisymmetrizing the two $\BB_2$ factors.
For example
\be
 \Li_{1,3}(-1/w,v w)\big|_{\BB_2\wedge \BB_2} =
 \big(\left[ v\wedge 1+v \right]\otimes \left[w\wedge 1+w \right]\big)
 -\big(\left[ w\wedge 1+w \right]\otimes \left[v\wedge 1+v \right]\big).
\ee This operation was famously used in ref.~\cite{Goncharov:2010jf}, where it was observed that the symbol of the 6-point remainder function
(in four-dimensional kinematics) had vanishing $\BB_2\wedge \BB_2$ component. This enabled its expression in terms of classical polylogs,
following a conjecture of Goncharov. Our situation is different: the symbols we find have non-vanishing $\BB_2\wedge \BB_2$ component.

\subsection{Integration method}
\label{sec:method}

Let us now detail the steps we have followed to obtain the form (\ref{mainresult2loop}).

First, we have computed the symbol of $f^{2(2)}_{8,1}$, starting from the integro-differential equation (\ref{diff2loop}) and using an automated
implementation of the algorithm described in Appendix A of ref.~\cite{arXiv:1105.5606}. The symbol turns out to have 68 terms with entries
$(v,1+v,w,1+w,1-v w)$.

Then we focused on its $\BB_2\wedge\BB_2$ projection; this turned out to be simply
\be
 f^{2(2)}_{8,1}\big|_{\BB_2\wedge \BB_2} \simeq -2(\Li_2(-v)-\Li_2(-w)) \wedge \Li_2(1-v w)\,. \label{B2B2symbol}
\ee
To match it, we constructed an Ansatz with two-arguments\footnote{One could generalize the Ansatz by adding higher-depth polylogs,
\emph{i.e.} $\Li_{1,1,2}$, but we did not consider this.}
 polylogs $\Li_{1,3}(a,b), \Li_{2,2}(a,b)$ and $\Li_{3,1}(a,b)$.
The arguments $a,b$ are rational functions of $v, w$ subject to the following constraints:
\begin{itemize}
\item[1.] The combinations $a,b,1-a,1-b$ and $a-b$ must factor (up to overall sign) into products of $v,1+v,w,1+w,1-vw$.
\item[2.] The quantities $1-a$ and $1-a b$ must remain non-negative for positive $v$ and $w$.
\end{itemize}
Constraint 2 ensures reality and single-valuedness in the Euclidean region, as can be seen
from the integral representation in \mbox{appendix \ref{app:polylogs}}.

Physically, these two conditions imply that the functions only have branch cuts at physical locations, not only on the first Riemann sheet (constraint 2) but on any Riemann sheet (constraint 1).
Neither constraint is strictly necessary -- rather these are a wish list of desirable properties.
The answer could conceivably be a complicated combination of polylogs which individually
violate these conditions.  However, our first Ansatz should reflect optimism.

Arguments satisfying these conditions can be generated systematically and we found a large number ($\sim200$) of them,
for example $\Li_{1,3}(-1/v,v w)$ or $\Li_{1,3}(\frac{vw-1}{v(1+w)},-v)$.  The first question is whether the $\BB_2\wedge\BB_2$ projection (\ref{B2B2symbol}) can be reproduced by combinations of these functions.  This turned out to be possible in a very large number of ways, with all but 6 coefficients left undetermined. (If no solution had been found, we would have had to backtrack and relax some assumptions.)

The counting of the solutions is easy to understand.
Within the set of considered symbol entries, $\BB_2$ is 4-dimensional (it is generated by $\Li_2(-v),\Li_2(-w),\Li_2(1-vw)$ and
$\Li_2(\frac{vw-1}{v(1+w)})$). The space of possible $\BB_2\wedge \BB_2$ is thus 6-dimensional.

One faces the problem of selecting the ``best'' solution.
Depending on the choice made here, the number of terms required to match the ``simpler" part of the answer,
\emph{i.e.} $\Li_4$, $\Li_3\log$ etc. will be either very large or small.  How to find the solution which minimizes (or nearly minimizes) the number of these terms?  We found that a successful strategy was to impose a few more desirable properties on the functions in our Ansatz.
\begin{itemize}
\item[3.] All terms in their symbols should have at most one entry of the form $(1-vw)$.
\item[4.] The functions should vanish at $v\to 0$.
\end{itemize}
In addition we dropped all $\Li_{i,j}$ functions which led to vanishing $\BB_2\wedge \BB_2$ projections.

Constraint 3 is natural as it is a property of the desired symbol. Entries of this type open the door to entanglement of the $v$ and $w$ dependence and we want to keep this entanglement as minimal as possible.

Constraint 4 requires some explanation.  The (imprecise) idea is that the series expansion of the physical amplitude, which is governed by the
collinear OPE of ref.~\cite{Alday:2010ku}, should be ``nice''. By expressing it in terms of functions which have a simple expansion in a given
channel, say the $v$ channel, one hopes to obtain more concise expressions. (In keeping with this principle, we systematically favored
polylogarithms with arguments $v$ and $w$, as opposed \emph{e.g.} to $1/(1+v)$, throughout the rest of the computation.)

These constraints further reduce the size of the Ansatz down to 12:
\ba
&& \Li_{2,2}(-1/w,v w), \quad
 \Li_{2,2}(-v,1/v/w), \quad
 \Li_{2,2}(-v/(1+v),(1+v)/v/(1+w),\quad
 \nl &&
 \Li_{2,2}(1/(1+w),v(1+w)/(1+v)), \quad
 \mbox{and same with } \Li_{2,2}\mapsto \Li_{1,3} \mbox{ or } \Li_{3,1}\,.
\ea
Within this reduced Ansatz we still found 11 free parameters.
We selected the combination in eq.~(\ref{mainresult2loop}) on the grounds that it seemed to make the length of the remaining symbol (as measured by number of terms) particularly small.

After subtracting these we find that we get the correct $\BB_4$ projection of the symbol up to a simple $-4\Li_4(-v)$.
The most complex remaining part is then encoded in the $\BB_3\otimes \SS_1$ projection, which is guaranteed
to be representable by weight 3 functions times logarithms.  Since the logarithmic part sits in the $\SS_1$, the problem is reduced to integrating a weight 3 symbol. This is similar to the weight 4 case just considered, but is, evidently, simpler.

Once the weight 3 polylogs are accounted for, the following components of the symbol are dealt with successively, in decreasing order of complexity: $\BB_2\otimes \BB_2$, $\BB_2\otimes \SS_1^2$ and $\SS_1^4$.  These correspond directly to $(\Li_2)^2$, $\Li_2\log^2$ and
$\log^4$ functions, respectively.

This way we arrive at a function which has the correct symbol and the correct branch cut structure.
The (almost) final step is to fix the beyond-the-symbol ambiguities (terms proportional to $\zeta(2)$ or $\zeta(3)$) by checking that the function obeys the correct differential equation.  Actually, after imposing the collinear-soft constraints, one can show that there is a unique beyond-the-symbol ambiguity, proportional to the one-loop amplitude times $\zeta(2)$.  We could easily fix its coefficient by numerically integrating eq.~(\ref{diff2loop}).

We would like to stress that the whole procedure is essentially automated, reduced at every step to linear algebra. One faces a number of
choices in constructing Ansatzes, where human input may help optimize the outcome. However, we found that once we had identified the guiding
principles 1 -- 4  the choices were rather straightforward.

\section{The three-loop MHV octagon}\label{3loop}

\subsection{The three-loop octagon from $\bar Q$}

The $\bar Q$ equation allows us to seamlessly upgrade the two-loop NMHV decagon just found, to the three-loop MHV octagon. Applying
eqs.(\ref{d0slash1}) and (\ref{d1slash2}) to the NMHV decagon, which turns the R-invariants into $\bar Q$ of bosonic quantities, it is apparent
(to any loop order) that the contribution from edge 8 to the right-hand side of the $\bar Q$ equation (\ref{Qbar}) takes the form \be
 \bar Q R_{8,0} \big|_\textrm{edge 8} = F_1(v_1,v_2) \bar Q \log \frac{[4\,8]}{[2\,8]} + F_2(v_1,v_2) \bar Q \log \frac{[6\,8]}{[2\,8]}
 \label{QbarMHVedge8}
\ee
where $F_1$ and $F_2$ are (as we will see shortly, pure) transcendental functions of cross-ratios.
Since the arguments of the $F_i$ are manifestly cross-ratios it is simple to use dihedral symmetry to add together the four even edge contributions,
and the arguments of $\bar Q$ then also become cross-ratios:
\be
 \bar Q R_{8,0} = F_1(v_1,v_2) \bar Q \log\frac{(1+v_2)^2}{v_2} + \big(F_2(v_1,v_2)-F_2(1/v_1,1/v_2)\big) \bar Q \log v_2\,.
\label{Qbar80a}
\ee
This holds provided that $F_1(v_1,v_2)=F_1(1/v_1,1/v_2)$ is satisfied, which can be viewed as a nontrivial constraint on the NMHV decagon
similar to (\ref{nontrivial0}).
Finally, since $\bar Q$ is a first order differential operator with respect to even bosonic variables, in a last step one can replace it
in eq.~(\ref{Qbar80a}) by $\partial/\partial u_2$, thereby obtaining the derivatives of the octagon.

Proceeding as in the NMHV case and following the steps in section \ref{2loopNMHV}, we have obtained the symbol of the MHV octagon, found the
corresponding transcendental functions, and fixed all integration constants through numerical integration of \ref{Qbar80a}. We directly state the result:
\be R_{8,0}^{(3)} = \left[\big(f_{8,0}^{a(3)}(v,w)+(v\leftrightarrow \frac 1 v)\big)+(w\leftrightarrow \frac 1
w)+f_{8,0}^{b(3)}(v,w)\right]+(v\leftrightarrow w)+f_{8,0}^{c(3)}(v,w).\label{3loopoctagon}\ee The ``non-trivial'' part is contained solely in
$f_{8,0}^{a(3)}$:
\begin{small}\ba f_{8,0}^{a(3)}(x,y)&=&2~\Li_{2,2,2}(\frac 1 {1{+}x},1{+}x,\frac 1 {1{+}y})+2\Li_{1,2,2}(\frac 1 {1{+}x},1{+}x,\frac 1
{1{+}y})\log(1{+}x)+\Li_{2,4}(-\frac 1 x, x y)+\Li_{1,4}(-\frac 1 x,x y)\log x\nl &&-\Li_{2,2,2}(\frac 1 {1{+}x},1{+}x,1)-\Li_{2,2,2}(1,1,\frac
1 {1{+}y})-\Li_{1,2,2}(\frac 1 {1{+}x},1{+}x,1)\log(1{+}x).\label{mainresult3loop1}
\ea\end{small}%
The remaining functions involve only classical or lower-weight polylogs, with various symmetry properties:
\begin{small}\ba
 f_{8,0}^{b(3)}(x,y)&=&-2\log\frac{(1+y)^2} y\left[5\Li_5(-x)-\Li_4(-x)\log
x+7\zeta_4\log(1{+}x)+\frac 1 6 \zeta(2)\log^3 x +\frac 1 4 \zeta_2\log\frac{(1{+}x)^2} x \log^2 x\right]\nl
&&-\left[2\Li_{3,1}(-x,1)+2\Li_{2,2}(1,-x)+2\Li_4(-x)+6\Li_{3,1}(1,-x)-2\log x\Li_{2,1}(1,-x)+4\zeta_3\log(1{+}x)\right]\nl
&&\times\log(1{+}y)\log\frac{1{+}y} y+(\frac 1 {24} \log^4 x-\frac{31} 4 \zeta_4)(\frac 1 2\log^2 y-\zeta_2)\nl
f_{8,0}^{c(3)}(x,y)&=&2~\left[2\Li_3(-x)-\Li_2(-x)\log x-\frac 1 2\log^2x\log(1{+}x)+\frac 1 6 \log^3 x -\zeta_2\log\frac{1{+}x}x\right]\times
(x\leftrightarrow y)-\frac {67\pi^6} {1260}\nl
&&-\log\frac{(1{+}y)^2} y\log\frac{(1+x)^2} x\left[\frac 2 3 \log(1{+}y)\log(1+\frac 1 y)\log(1{+}x)\log(1+\frac 1 x)+\frac 1 {12}\log^2 x\log^2
y-\frac 9 2\zeta_4\right]. \nl\label{mainresult3loop} \ea\end{small}%
Equations (\ref{3loopoctagon})-(\ref{mainresult3loop}) are one of the main results of of this paper.

Proceeding as in section \ref{sec:upliftNMHV} we can uplift this formula by defining
\be
 S_{8,0}^{(3)} = \left[f_{8,0}^{a(3)}(v,w)+f_{8,0}^{a(3)}(\frac 1 v, \frac 1
w)+\frac12 f_{8,0}^{b(3)}(v,w)\right]+(v\leftrightarrow w)+\frac12 f_{8,0}^{c(3)}(v,w) + \delta_{8,0}^{(3)}(v,w)
\ee
where $\delta_{8,0}^{(3)}(v,w)$ is an undetermined function with the symmetry properties $\delta_{8,0}^{(3)}(v,w)=\delta_{8,0}^{(3)}(w,v)=-\delta_{8,0}^{(3)}(1/v,w)$.  It turns out that it can be chosen such that $S_{8,0}^{(3)}$ does not diverge in on-shell limits, and by computing the 10-point symbol we found that it could also be chosen such that only $\log^6$ and $\Li_2\log^4$ terms remain at 10-points after subtracting the uplifting.
Furthermore, all terms of the symbol of the remainder turn out to have three odd twistor entries and three even twistor entries.
However, since the 10-point MHV amplitude is yet insufficient to fix the complete $n$-point amplitude\footnote{
One would need $R_{12,0}^{(3)}$, which would be straightforward if the complete all-$n$ uplifting of two-loop NMHV amplitudes could be determined.},
we do not attach here the formula for $S_{8,0}^{(3)}$ but simply state a general conjecture in conclusion, leaving open the computation of the complete $n$-point 3-loop amplitude.

\subsection{Numerical comparison with two-loop and strong coupling}
\label{sec:numerics}

\begin{figure}\centering
\subfigure[$\bar R^{(3)}_{8,0}$]
    {
        \includegraphics[width=4.5cm]{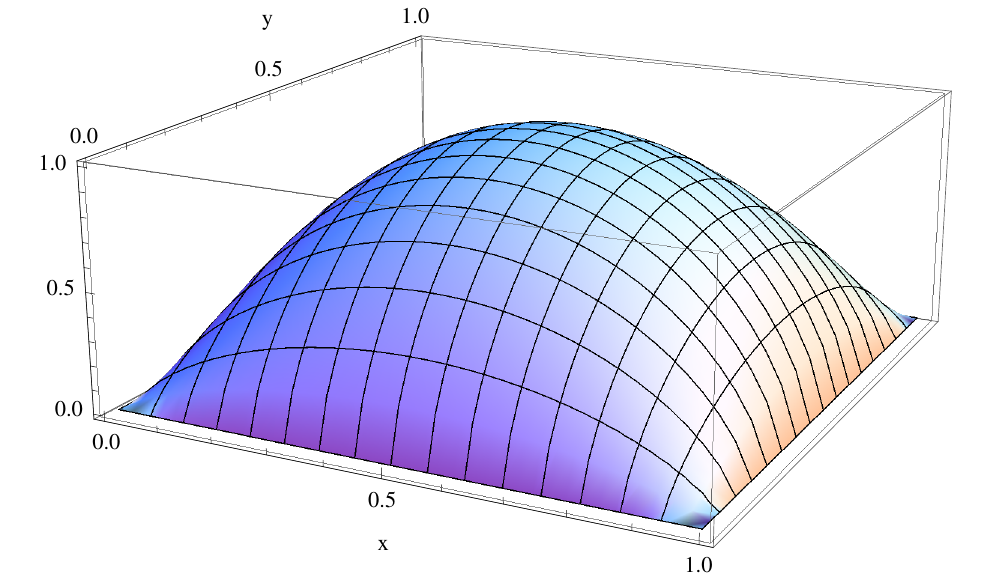}
        \label{fig:3loopa}
    }
    \subfigure[$\bar R^{(3)}_{8,0}-\bar R^{(2)}_{8,0}$]
    {
        \includegraphics[width=4.5cm]{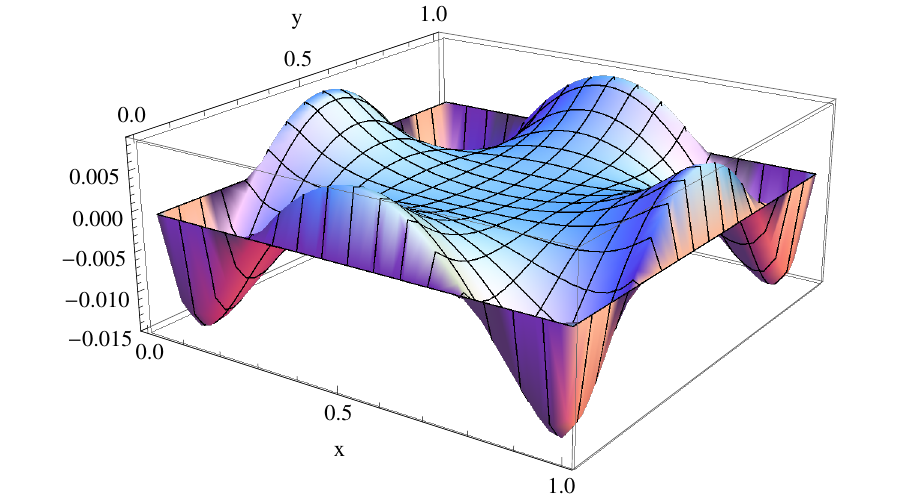}
        \label{fig:3loopb}
    }
    \subfigure[$\bar R^{(2)}_{8,0}-\bar R^\textrm{strong}_{8,0}$]
    {
        \includegraphics[width=4.5cm]{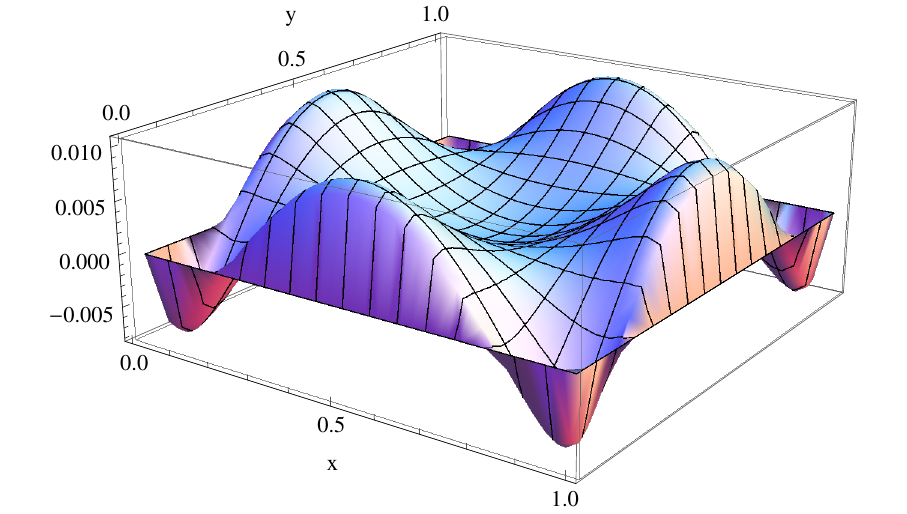}
        \label{fig:3loopc}
    }
    \caption{The octagon remainder function in the Euclidean region: (a) the normalized 3-loop function; (b) the difference between 3-loop and 2-loop normalized functions; (c) the difference between 2-loop and the strong coupling normalized functions.}
    \label{fig:3loop}
\end{figure}
\begin{figure}\centering
\subfigure
{\includegraphics[width=7cm]{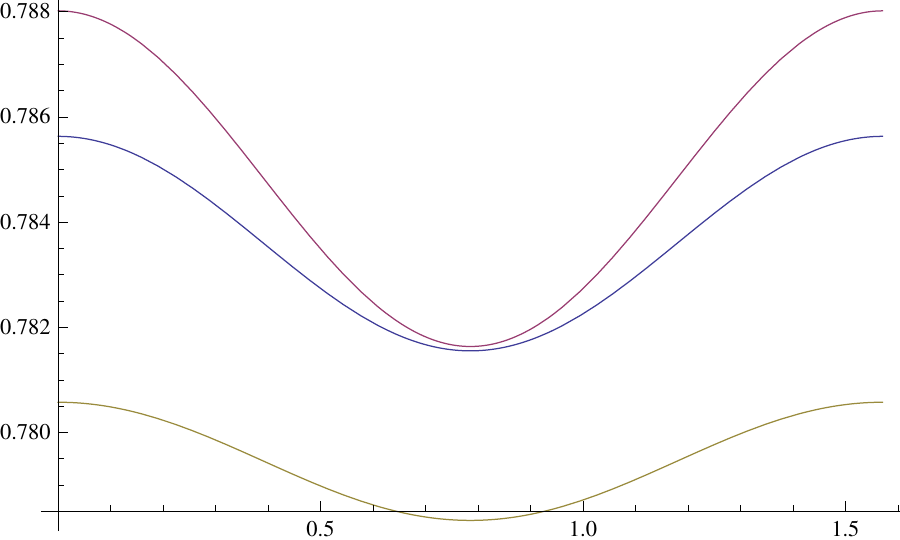}
}
\subfigure
{
\includegraphics[width=7cm]{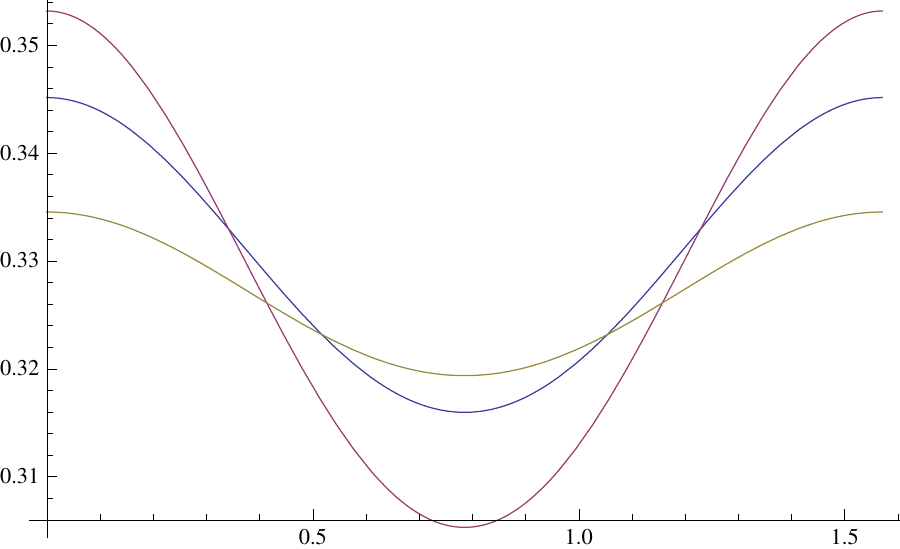}
}
\caption{Plot of $\bar R^{(2)}_{8,0}$ (blue), $\bar R^{(3)}_{8,0}$ (red) and $\bar R^\textrm{strong}_{8,0}$ (yellow) as a function of $\phi$, at $|m|=0.2$ (left) and $|m|=0.45$ (right).}~\label{fig:phidep}
\end{figure}

We now present a numerical comparison of the three-loop octagon with two-loop and strong coupling results.  In Fig.~\ref{fig:3loopa} we plot the three-loop remainder function in the Euclidean region --- the region defined by the two cross-ratios $v,w$ being positive, normalized
by its value at the symmetrical point, $\bar R_{8,0}^{(i)}\equiv R_{8,0}^{(i)}/R_{8,0}^{(i)}(v=w=1)$, with the value at the symmetrical point being
$R_{8,0}^{(3)}(v=w=1)\approx 1.837584$.

Despite appearances $R_{8,0}^{(3)}$ does not have a definite sign --- it goes slightly negative near the corners of the Euclidean region.
This can be verified analytically using the expansion around collinear limits given in the next subsection.

A surprising feature is that the shape of the three-loop amplitude is approximately the same as at two-loops, at least in the Euclidean region.
The residual (difference between the three-loop and two-loop amplitudes, both normalized by their values at the symmetrical point) is shown in Fig.~\ref{fig:3loopb}
and is seen to be at most ~1$\%$ of the maximal value, in absolute values.

This is very similar to what was observed in refs.~\cite{Brandhuber:2009da,arXiv:1006.4127}, with strong coupling plotted using the formulae of ref.~\cite{Alday:2009yn}:
\be
 \bar R_{8,0}^\textrm{strong} = -\frac12 \log(1+v)\log(1+1/w) + \int_{-\infty}^\infty dt\frac{|m|\sinh t}{\tanh(2t+2i\phi)}\log(1+e^{-2\pi|m|\cosh t})
\ee
where $|m|$ and $\phi$ are defined through $v=e^{2\pi|m|\cos \phi}$ and $w=e^{2\pi |m|\sin\phi}$. Note that the constant in~\cite{Alday:2009yn} is gone because of our definition of $R$ in \Roneone kinematics, $R_{6,0}=1$ to all loops.

Near the boundaries of the Euclidean region, where the amplitude is small, the relative error is significant as visible in Fig.~\ref{fig:3loopc}.
To see the comparison more directly, we can plot the three octagon remainder functions $\bar R^{(2)}_{8,0}$, $\bar R^{(3)}_{8,0}$ and $\bar R^\textrm{strong}_{8,0}$  as a function of $\phi$, for given values of $m$; see Fig.~(\ref{fig:phidep}). A similar comparison between 2-loop and strong coupling functions was plotted in Figure 2 of ref.~\cite{arXiv:1006.4127}. As we can see very clearly, taking into account that its overall sign is opposite, adding the three-loop correction indeed makes the shape of the function closer to the strong coupling one.

The Euclidean region is of course not the only physically interesting region.  Although the functions may look numerically similar in the Euclidean region, they become numerically very different after continuation to Lorentzian regions.  For example, in the ``Regge'' limit discussed below,
$R_{8,0}^{(2)}$ vanishes while $R_{8,0}^{(3)}$ does not.  Such a numerical insensitivity to the interaction strength in the Euclidean region was observed previously in other contexts (see for instance \cite{Teaney:2006nc}), and it would be very interesting to have an explanation for it.

\subsection{Expansion around collinear limits}

In ref.~\cite{Alday:2010ku} a systematic expansion around collinear limits was developed. This expansion takes the form of an OPE, and in the
present case allows to expand in powers of $v$ at fixed $w$.  Of particular interest are the logarithmically enhanced terms $\sim \log v$
(multiplying arbitrary powers of $v$), also called OPE discontinuities,
since at $\ell$-loops these can typically be predicted in terms of known anomalous dimensions times
lower-loop form factors.

Using the formulas in appendix \ref{app:polylogs} we have expanded the preceding analytic result around $v=0$ and verified
that the double logarithmic terms ($\sim \log^2 v$) reproduce the predictions of ref.~\cite{arXiv:1010.5009}. The single logarithmic terms are
more lengthy, and a number of these terms display an interesting feature: they involve powers of $w^i$ or $1/w^i$ multiplying transcendental
functions of $w$.  These terms, which we will refer to as ``mixing terms'', are given by
\begin{small}\be R^{(3)}_{8,0}=\sum^\infty_{n=1}v^n \left(
c^{(1)}_n[2~\Li_2(-w){+}\log w\log(1{+}w){+}\log\frac{w}{1{+}w}\log \frac{v}{1{+}v}]{+}c^{(2)}_n\log\frac{w}{1{+}w}\right)_{\textrm{reg}}
+(w\leftrightarrow \frac 1 w){+}\ldots\,,~\label{expansion}\ee\end{small}%
where $\ldots$ contains only transcendental functions of $w$ with pure prefactors (independent of $w$); ``reg" means to remove a polynomial in
$w$ and $\log w$ such that the expression vanishes as $w\rightarrow \infty$, and we add the other term given by $w\leftrightarrow 1/w$.
With the definition
$S_{i,j}\equiv\sum^j_{k=1} \frac 1 {k^i}$, $c^{(1)}_n$ and $c^{(2)}_n$ are given by
\begin{small} \ba c^{(1)}_n(w)&=&\sum^n_{m=1}\frac{(-)^{n{-}m{+}1}}{n\,m}\left(4S_{2,n}{-}4S_{2,m}{-}\frac2{n^2}{+}\frac2{m^2}\right)w^m{-}\frac{4 w^n}{n^4}\,,\nl
c^{(2)}_n(w)&=&\sum^n_{m=1}\frac{(-)^{n{-}m}} {n\,m}\left(8S_{3,n}{-}8S_{3,m}{+}(4S_{2,n}{-}4S_{2,m})(\frac 1 n{+}\frac 1 m){-}\frac
6{n^3}{+}\frac 6 {m^3}{-}\frac 2 {n^2\,m}{+}\frac
2{m^2\,n}\right)w^m{+}\frac{16 w^n}{n^5}\,.\nl~\label{c1c2}\ea \end{small}%

At the lowest nontrivial order, two-loop, the OPE discontinuities of the remainder function are accounted for by exchange of single $F_{+-}$ excitations and their derivatives.  We have tried to see if the next-order result could be accounted for by correcting the form factors and using the (known) two-loop energies for these excitations. As may have been anticipated, there is an obstruction: given that the two-loop MHV amplitude has no mixing term, this procedure will never produce any mixing term!

For twist $2n\geq 4$ this obstruction is not surprising, since states with two $F_{+-}$ excitations are expected to contribute.
However, for twist $2n=2$ such states do not exist, so we interpret the above obstruction as evidence that the $F_{+-}$ excitations decays into two
twist-one fermionic excitations, as suggested in ref.~\cite{arXiv:1010.5009}.

To understand better the $1/w$-enhanced terms, we consider the analytic continuation described in ref.~\cite{Bartels:2010tx} in the context of
the multi-Regge-limit of the hexagon (in four-dimensional kinematics): for fixed (small) $v$ we take $w$ around its branch cut at $w=-1$, and
come back  to the origin.  Note that eq.~(\ref{expansion}) vanishes as $w\to 0$, but no longer after going through the cut. In fact, it directly
follows from eq.~(\ref{c1c2}) that the term $\sim v^n$ diverges like $1/w^n$ after the continuation,
so the limit $v\to 0$ with $v/w$ fixed gives an interesting
function, similar to ref.~\cite{Bartels:2010tx}: \be \lim_{\substack{v\to 0\\v/w\textrm{ fixed}}}{R_{8,0}^{(3)}}'= 2\pi i \left(4\log
\frac{v}{w}\Li_4(\frac{v}{w})-16~\Li_5(\frac{v}{w}) \right) + \mathcal{O}(|vw|) \ee where the prime refers to the fact that we are on the second
Riemann sheet. By analogy with~\cite{Bartels:2010tx} we call this limit a Regge limit. By the preceding discussion, the fact that this limit is nontrivial (in
contradistinction with the earlier proposal in~\cite{Heslop:2011hv}) seems here to be intimately tied to the decay of the $F_{+-}$ excitation. We hope to
explore this connection further in future work.

\section{Conclusion}

In the \Roneone kinematics, the S-matrix in planar $\mathcal{N}=4$ SYM simplifies significantly and possesses remarkable structures. In this
paper, we have explicitly computed amplitudes/Wilson loops in two-dimensional kinematics, by exploiting the $\bar Q$ symmetry and constraints from collinear limits.
Eqs.~(\ref{8ptN2MHV}), (\ref{mainresult2loop}),  (\ref{generalnmhv2}) and (\ref{3loopoctagon})-(\ref{mainresult3loop}) are our main results: we have
obtained, for the first time, compact analytic formula for three classes of octagons: one-loop N${}^2$MHV, two-loop NMHV and three-loop MHV
respectively. Furthermore, we provide strong evidence that, by using the collinear uplifting, the octagon results can be straightforwardly
generalized to higher-point ones; in particular, we have identified the fundamental eight-point building block which can be uplifted to all
multiplicities. As a lower-loop example of the uplifting formalism, we have also presented a new explicit, manifestly dual conformal invariant
expression for $n$-point one-loop NMHV amplitude.

We have obtained these results by combining two complementary methods: we went to higher loops using $\bar Q$ equation, and to higher points
with the help of collinear uplifting. It is important to stress that, although nontrivial, our final analytic formulaeare still rather compact and manageable. Finding analytic expressions for higher-loop amplitudes in \Roneone kinematics will be harder 
but at least one more loop would seem technically feasible. It is straightforward to use $\bar Q$ equation to go to four loops, if two-loop N${}^2$MHV amplitudes can be obtained somehow; on the other hand, combining our equations with constraints from OPE expansion (especially of the type obtained in~\cite{Basso:2013vsa}) might be powerful enough to determine the four-loop octagon.  More generally, it would be interesting to explore the connections between our formalism and the OPE expansion of Wilson loops:  our results provide valuable data for better understanding single and multi particle excitations, and the OPE picture should shed more lights on the integrability underlying our formalism.

Without explicitly going to more loops or more legs, one can make predictions about the structures at all loops. Based on our $k{+}\ell=3$ results, we propose the following three conjectures:
\begin{itemize}
\item For octagons in \Roneone kinematics to all loops, only six letters can appear in the symbol: $v$,$w$,$1{+}v$,$1{+}w$,$v{-}w$ and $1{-}vw$, all of which are already seen at three loops.
\item The last entry of the symbol for MHV and NMHV octagons can only be $v$,$w$,$1{+}v$ or $1{+}w$.
\item The $\ell$-loop N${}^k$ MHV amplitude in \Roneone amplitudes can be obtained by collinear-uplifting the octagon, dodecagon, etc.
up to a $4(\ell{+}k)$-gon.  That is, by uplifting basic building blocks of the type $S_{8}$, $S_{12}$, \ldots,$S_{4(\ell{+}k)}$.
Furthermore, the depth of the transcendental functions entering $S^{(\ell)}_{4m}$ should be at most $\ell+2-m$.\footnote{We are unsure about the standard definition of ``depth'' for multiple polylogarithms, but we define it as the maximal number of antisymmetric pairs which can be extracted from the symbol of the function. For example, $\log^4x$ has depth 0 while $\Li_2(x)^2$ has depth 2.}
\end{itemize}
A special case of the third conjecture is that the yet missing purely logarithmic terms in NMHV two-loop for $n>8$, or the purely logarithmic times dilogarithmic terms for MHV three-loop with $n>8$, should be generated by ``uplifting'' $S_{12}$ objects.  This conjecture is a refinement of that in ref.~\cite{Goddard:2012cx}, where we have included the notion of depth to stress that the functions $S_{4m}$ should become simpler and simpler as $m$ increases.  Also, note that instead of allowing for $S_{10}$'s, we instead allow for $S_{4m}$'s which do not vanish in collinear limits (see eq.~(\ref{generalnmhv2})). It would be desirable to obtain a closed form for $S_{12}$ objects, and completely determine the two-loop NMHV and three-loop MHV $n$-gons.

It is very interesting to observe that the structure of multi-loop amplitudes in \Roneone bears similarity to that of correlation functions. For example, multi-loop integrals for the four-point correlation function receive contributions from``mixing'' terms,  e.g. $x{-}\bar{x}$ and $1{-}x\bar{x}$ at three and four loops, where $x$,$\bar{x}$ are related to the two cross-ratios~\cite{Drummond:2013nda}. It would be nice to study if there are relations between (integrals of) on-shell amplitudes/Wilson loops and off-shell correlators.

It is also worth noticing that the three new octagons we obtained, with $k{+}\ell{=}3$, share a curious factorization property:
their non-trivial parts ``factorize'' into the four-mass box and simple functions of corresponding weights from one to three loops. We leave the detailed discussion to~Appendix.~\ref{app:property}, and an interesting open question is if similar structures appear at higher loops.
We also noted some numerical ``accidents'' in section \ref{sec:numerics} --- the two-loop, three-loop and strong coupling remainder functions have similar shapes at the percent level, when restricted to the Euclidean region.

Finally, we comment on perhaps the most remarkable aspect of the $\bar Q$ formalism, which is that the $\bar Q$ equations can be solved at all. This is ultimately a consequence of the derivation in \cite{CaronHuot:2011kk}, but in practice at $\ell$-loop order this imposes nontrivial constraints on \emph{lower-loop} amplitudes, as we discuss around (\ref{mainresult2loop}). These constraints relate, for instance, various collinear limits of the one-loop N$^2$MHV decagon to each other.
The fact that the decagon can be consistently ``integrated'' to give the two-loop NMHV octagon is very nontrivial. Similarly, the fact that the octagons can be ``integrated'' to give the (trivial) hexagons is also nontrivial.  It would be very interesting to formulate precisely and explore these constraints.

\section{Acknowledgement}
SCH gratefully acknowledges support from the Marvin L.~Goldberger Membership and from the National Science Foundation under grant PHY-0969448. SH wishes to thank Jan Plefka for an invitation to visit Humboldt University of Berlin during the completion of the work.

\begin{appendix}

\pagebreak
\section{BCFW recursion relations in two dimensions}\label{app:tree}

The NMHV amplitudes can be easily obtained from their well-known four-dimensional expression,
\be
 R^\textrm{$D$=4,tree}_{n,1}=\sum_{1\leq i<j\leq n} [*\,i\,i{+}1\,j\,j{+}1]\,,
\ee
by exploiting independence of this formula upon the choice of reference momentum twistor $*$. Note here $[*\,i\,i{+}1\,j\,j{+}1]$ is the four-dimensional R-invariant depending on five labels, $[a\,b\,c\,d\,e]:=\frac{\delta^{0\|4}(\l a\,b\,c\,d\r\eta_e+\mbox{cyclic})}{\l a\,b\,c\,d\r ...\l e\,a\,b\,c\r}$;
indeed for the following choice, the two-dimensional limit is non-singular for each term:
\be
 \ZZ_*=\ZZ_1+\ZZ_2,\qquad [*\,i\,i{+}1\,j\,j{+}1] \to -(1\,i\,j)[2\,i{+}1\,j{+}1], \quad \mbox{for $i,j$ even}\,.
\ee
There are similar expression when $i$ or $j$ are odd, interchanging  $i$ and $i{+}1$ as required to put the odd argument into the $(1,\ldots)$ parenthesis and the even argument into the $[2,\ldots]$ bracket.
In this way the four-dimensional formula is reduced directly to
\ba
  R^\textrm{tree}_{n,1}&=&\sum_{3\leq i<j\leq n{-}1} (1\,i\,j) \big( [2\,i{+}2\,j{-}1]+[2\,i{-}1\,j{+}1]-[2\,i{+}1\,j{+}1]-[2\,i{-}1\,j{-}1]  \big)\nl
  &=&\sum_{3\leq i<j\leq n{-}1} (1\,i\,j) \big( [i{-}1\,i{+}1\,j{-}1]-[i{-}1\,i{+}1\,j{+}1]\big)
 \label{nmhvtree2d}
\ea where the second line follows from the first via the four-term identity, $(i\,j\,k)+(j\,k\,l)+(k\,l\,i)+(l\,i\,j)=0$.

To describe the N${}^k{-}2$MHV tree amplitudes it is better to construct them recursively staying in two dimensions.
We achieve this by means of an adapted BCFW deformation,
\be
 \ZZ_{n{-}1}\to \hat \ZZ_{n{-}1}(w) := \ZZ_{n{-}1} + w \ZZ_{1}\,,
\ee
which respects the two-dimensional kinematics.
Following the BCFW argument, we have to understand the poles of the amplitude as a function of $w$.
Since the factorization properties on poles are universal, it suffices to understand the poles of eq.~(\ref{nmhvtree2d}). Thus in a sense we will bootstrap our way up from NMHV to general N${}^k{-}2$MHV.

The simplest factorization occurs as $w\to\infty$, where the segment $n$ becomes soft.
From the definition of the R-invariants it can be seen that
$(n{-}1\,1\,i)\to 0$ ($\sim1/w$) in that limit
and that $\ZZ_{n{-}1}\to \ZZ_{1}$ in all other terms.  From eq.~(\ref{nmhvtree2d}) it is then easy to see that
\be
 \left.\hat R^\textrm{tree}_{n,1}\right|_{w\to\infty} \to R^\textrm{tree}_{n{-}2,1}(1,\ldots,n{-}2).
\ee
This has the expected factorized form. 
By universality of factorization, we conclude that this expression will remain true for all $k$.

\begin{figure}\centering
\includegraphics[height=4cm]{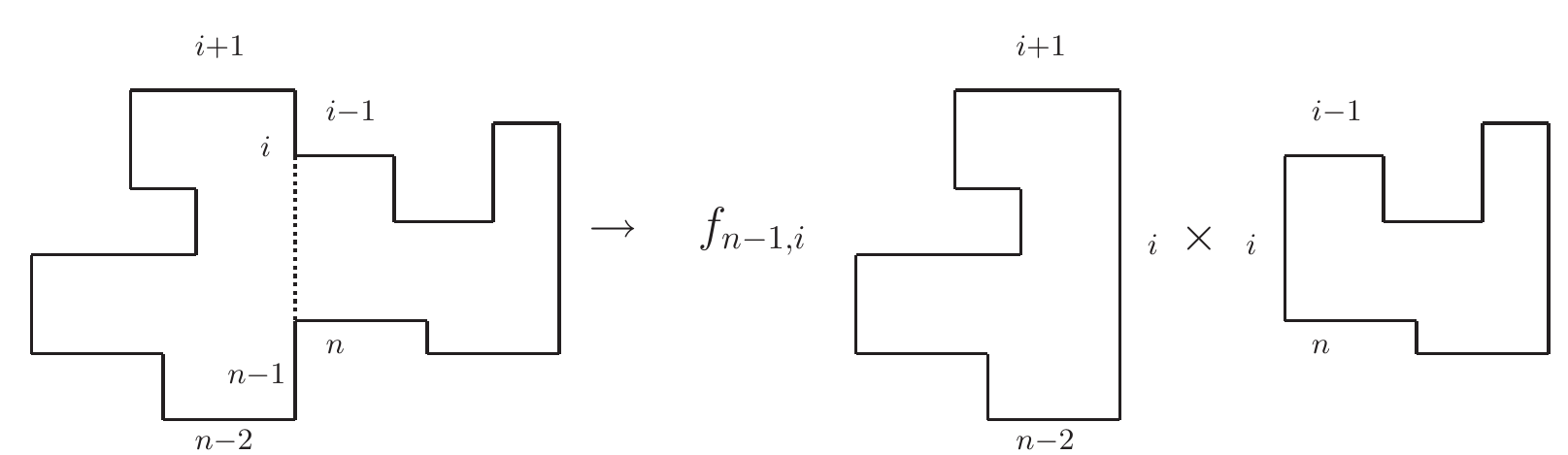}
\caption{Wilson loops and factorizations in two-dimensional kinematics: the zigzag null polygon is parametrized by the coordinates $(x^+, x^-)$  (rotated by 45 degrees). When $\l \widehat{n{-}1},i\r=0$,  the $x^-$ (horizontal) coordinates of side $i$ and
side $n{-}1$ coincide, and the amplitude becomes the product of the two lower-point amplitudes, with a prefactor
$f_{n{-}1,i}:=(n{-}1\,1\,i)\big( [n{-}2\,n\,i{+}1]-[n{-}2\,n\,i{-}1]\big)$.} \label{fig:BCFW}
\end{figure}

The generic pole of $\hat R_n(w)$ occurs when $\l \widehat{n{-}1},i\r=0$ for each odd $i$, that is, $w=\frac{\l n{-}1\,i\r}{\l 1\,i\r}$. On
these poles we have \be \left. R^\textrm{tree}_n \right|_{\l n{-}1\,i\r\to0} \to (n{-}1\,*\,i)\big( [n{-}2\,n\,i{+}1]-[n{-}2\,n\,i{-}1]\big)\,.
\ee In the N${}^k$MHV amplitude, this universal prefactor will simply get multiplied by a product of two MHV-stripped super amplitudes, which
has a simple geometric interpretation as ``sliding'' one-side of the polygon. This is depicted in Fig.~\ref{fig:BCFW}.

Since these are all the singularities as a function of $w$, the BCFW argument immediately establishes the recursion formula, eq.~(\ref{BCFW}).
As an illustration, the 10-point N${}^2$MHV amplitude, which we used in section~\ref{N2MHV}, easily follows from the recursion, \ba
 R^\textrm{tree}_{10,2}&=&(1\,3\,5)(1\,5\,7)[2\,4\,6][2\,6\,10]+(1\,3\,5)(1\,7\,9)[2\,4\,8][2\,8\,10]+(1\,5\,7)(1\,7\,9)[2\,6\,8][2\,8\,10]-\nl&&
(1\,7\,9)(3\,5\,7)[2\,6\,8][2\,8\,10]+(1\,7\,9)(3\,5\,7)[2\,8\,10][4\,6\,8]-(3\,5\,7)(3\,7\,9)[2\,8\,10][4\,6\,8]-\nl&&
(1\,3\,5)(5\,7\,9)[2\,4\,10][4\,8\,10]+(3\,5\,7)(3\,7\,9)[4\,6\,8][4\,8\,10]+(1\,3\,5)(5\,7\,9)[2\,4\,10][6\,8\,10]\,.\nl \label{n2mhvtree}
\ea  We have verified the recursion relations arithmetically (by which we will mean, numerically for a random set of integer-valued kinematics) against the two-dimensional limit of the amplitudes given by the \emph{BCFW} package \cite{arXiv:1011.2447}.

\section{Remainder for one-loop N${}^2$MHV decagon}\label{app:r10}

The remainder, $r_{10,2}^{(1)}$, after we subtract contributions from $S_8$, has no rational prefactors. It depends on all the 10 external
points, for which supersymmetrically there are $6\times6=36$ independent R-invariants: in addition to the $10{+}10{+}5$=25 R-invariants in the
uplifting above, we can include e.g. the length-10 cyclic group generated by $(1\,3\,5\,7)[10\,2\,4][4\,6\,8]$, and the tree amplitude,
$R_{10,2}^\textrm{tree}$ (see eq.~\ref{n2mhvtree}), to form a basis. (In practice the different R-invariants can be easily separated from each
other, by evaluating the amplitude on different helicities.) We find \ba r^{(1)}_{10,2}&=&(1\,3\,5\,7)[4\,6\,8\,10](g_2{-}f_1){+}\mbox{(9
cyclic)}{+}(1\,3\,5\,7)[2\,4\,6\,8](g_1{+}g_2{+}f_2){+}\mbox{(9 cyclic)}{+}\nl&& (1\,3\,5\,7)[2\,6\,8\,10](g_1{+}g_2{+}f_3){+}\mbox{(4
cyclic)}{+}(1\,3\,5\,7)[10\,2\,4][4\,6\,8](g_1{+}g_2{-}f_4){+}\mbox{(9 cyclic)}{+}\nl&& R_{10,2}^\textrm{tree}(3
g_1{+}g_2+\frac{\pi^2}6)\,,\label{r10} \ea where we have cyclically symmetric functions $g_1, g_2$, and non-symmetric functions $f_1,...,f_4$,
\begin{small}
\ba g_1&=&\frac 12\sum^{10}_{i=1}\log(v_i)\log(v_{i{+}1}),\quad g_2=-\sum^5_{i=1}\log(v_i)\log(v_{i{+}5}),\nl
f_1&=&L(v_1){+}L(v_4){+}\sum^5_{i=1}\log(v_{i{-}1})\log(v_i){+}\log(v_1)\log(v_6){+}\log(v_4)\log(v_9){-}\log(v_5)\log(v_8){-}\log(v_7)\log(v_{10}){-}\frac{\pi^2}{12},\nl
f_2&=&L(v_1){+}L(v_2){-}\log(v_5 v_7)\log(v_6 v_8){-}\log(v_6)\log(v_7){+}2\left(\log(v_1)\log
(v_6){+}\log(v_2)\log(v_7)\right){-}\frac{\pi^2}{12},\nl f_3&=&L(v_1){+}L(v_6){-}\log(v_2)\log (v_5){-}\log (v_7)\log
(v_{10}){+}2\log(v_1)\log(v_6){-}\frac{\pi^2}{12},\nl\nl f_4&=&2L(v_1){-}\log(v_5
v_7)\log(v_6){+}2\log(v_1)\log(v_6){+}\log(v_2)\log(v_7){+}\log(v_5)\log(v_{10})\,.\ea\end{small} Here only 4 cross-ratios, out of
$v_1,...,v_{10}$, are independent. We have checked that eq.~(\ref{10ptN2MHV}) gives the correct $k$-preserving and $k$-decreasing collinear
limits.

\section{Evaluation of multiple polylogs}\label{app:polylogs}

In the main text we have used multiple polylogs, depending on several variables, defined by
\be
 \Li_{i_1,i_2,\ldots,i_k}(x_1,x_2,\ldots,x_k) := \sum_{a_1>a_2>\ldots>a_k\geq 1} \frac{x_1^{a_1}}{a_1^{i_1}}\frac{x_2^{a_2}}{a_2^{i_2}} \cdots \frac{x_k^{a_k}}{a_k^{i_k}}.
\ee
In this appendix we give integral representations for these functions, which we have found convenient both for numerical evaluation and for understanding their analytic properties.

We begin by recalling an integral representation for the one-index polylogs, or classical polylogs,
which can be obtained by integrating the ``seed''
$\Li_0(x)=x/(1-x)$ against logarithms
\be
 \Li_i(x) = \int_0^1 \frac{xdt}{(1-xt)} \frac{(-1)^{i{-}1}\log^{i{-}1} t}{(i{-}1)!}.
\ee
This can be proved easily by series-expanding in $x$ under the integration sign.

In the two-index case one can similarly integrate the ``seed'' $\Li_{0,0}(x,y)= x^2y/(1-x)/(1-xy)$ against two logarithms to obtain, for $i,j\geq 1$,
\ba
 \Li_{i,j}(x,y) &=& \int_0^1 \frac{x^2yt_1dt_1 dt_2}{(1{-}t_1x)(1{-}t_1t_2xy)}
  \frac{(-1)^{i{+}j}\log^{i{-}1} t_1 \log^{j{-}1} t_2}{(i{-}1)!(j{-}1)!} \nl
  &=& \int_0^1 \frac{xdt}{1-xt} \frac{(-1)^{i{-}1}\log^{i{-}1} t}{(i{-}1)!} \Li_j(xyt).
\ea
The validity of the first form can be proved easily by series-expanding in $x$ and $y$. We have found the second form,
which follows by integrating $t_2$ explicitly, particularly efficient for numerical analysis.

Applying the same procedure to three-indices objects similarly produces a three-fold integration, but again the last integral
can be done explicitly in terms of classical polylogarithms.
It is actually possible to perform one further integral, by keeping $t_1t_2$ fixed, and obtain a one-fold representation, which we record here
\ba
\Li_{i,j,k}(x,y,z)
  &=& \int_0^1 \frac{x^2yt_1dt_1 dt_2}{(1{-}t_1x)(1{-}t_1t_2xy)}
  \frac{(-1)^{i{+}j}\log^{i{-}1} t_1 \log^{j{-}1} t_2}{(i{-}1)!(j{-}1)!} \Li_k(xyzt_1t_2)
 \nl &=& (-1)^j \int_0^1 \frac{xy dt}{1-xyt} \Li_k(xyzt)
 \times \left( \sum_{k=0}^{i{-}1} \log^k t~\Li_{i{+}j{-}k{-}1}(xt) \frac{(-1)^k (i{+}j{-}k{-}2)!}{k!(i{-}k{-}1)!(j{-}1)!}
 \right. \nl && \hspace{2cm} \left.
-\sum_{k=0}^{j{-}1} \log^k t~\Li_{i{+}j{-}k{-}1}(x)\frac{(i{+}j{-}k{-}2)!}{k!(i{-}1)!(j{-}k{-}1)!} \right).
\ea
This can be used for efficient numerical evaluation.

\section{A curious factorization property for $k+\ell=3$ octagons\label{app:property}}
Here we discuss a factorization property of the three new octagon functions we obtained in the paper. Although with different loop orders, they
all lie on the same ``line of constant complexity'' $k+\ell=3$. To compare these functions, it is natural first to concentrate on the terms with
singularities at $1-v w=0$, which we will call the ''mixing'' part of the result.  Note that such terms are directly responsible for the multi-index polylogarithms and are absent at lower loop order; they were also absent from the Ansatz made in ref.\cite{Heslop:2011hv}.

According to eq.~(\ref{8ptN2MHV}), the nontrivial terms in the N${}^2$MHV octagon, e.g. the terms with a pole at $1-vw=0$, are \be
 R_{8,2}^{(1)} = -2 f_{2\,4\,6\,8} \times \frac{1}{1-vw}\times (1\,3\,5\,7)[2\,4\,6\,8]  + \textrm{non-mixing}\,,  \label{mixingpartR82}
\ee where \be \tilde f_{2\,4\,6\,8} = L(v)+L(w) + \frac12\big(\log v\log w-\log(1+v)\log w-\log v\log(1+w)\big)\,. \ee is the four-mass-box
function we discussed before. Thus eq.~(\ref{mixingpartR82}) states that the mixing part of $R_{8,2}^{(1)}$ comes entirely from
the four-mass-box integral, which result would follow easily from a Feynman diagram analysis.

We can similarly look at the mixing part of $R_{8,1}^{(2)}$. Here the entry $1-vw$ appears only in the third slot of the symbol, so it can be pulled out
by taking the $(2,1,1)$ component of the co-product (see ref.~\cite{Duhr:2012fh} and references therein).
Using (\ref{mainresult2loop}) we find \ba
 \Delta_{2,1,1} R_{8,1}^{(2)}& =& -2f_{2\,4\,6\,8} \otimes (1-v w) \otimes \log(vw)\left[ (1\,3\,5)[8\,2\,4]+(3\,5\,7)[2\,4\,6]+(5\,7\,1)[4\,6\,8]+(7\,1\,3)[6\,8\,2]\right]\,.\nl
&&+\mbox{non-mixing.} \label{mixingpart2loop}\ea Finally, $1-vw$ also appears only in the third slot of the symbol of the 3-loop octagon, and
the mixing part can thus be extracted from eq.(\ref{mainresult3loop}) by taking the $(2,1,3)$ component,\be
 \Delta_{2,1,3} R_{8,0}^{(3)} = -2f_{2\,4\,6\,8} \otimes (1-vw) \otimes \SS[f_3] + \mbox{non-mixing}
\ee\label{mixingpart3loop} where \ba
 f_3&=& 4\Li_3(1/(1{+}v))-4\Li_3(v/(1{+}v))+(2L(-v)-2L(-w))\log \frac{(1+v)^2}{v} +(v\leftrightarrow w)
 \nl && -\log(v w)\left[\log(1{+}v)\log(1{+}w)+\log(1+\frac{1}{v})\log(1+\frac{1}{w})\right]+\frac23\log^3(vw)\,.
\ea

It is striking that in all cases it is precisely the four-mass-box function which appears to the left of $(1-vw)$. It would be interesting to
understand why that is the case.

\end{appendix}

\end{document}